\documentclass[proceedings]{JHEP37}
\usepackage{eurosym}
\usepackage{amssymb}
\usepackage{amsfonts,cite}
\usepackage{amsmath}
\usepackage{graphicx}
\usepackage{multirow}
\usepackage{xcolor}
\usepackage{epsfig}

\newbox\mybox

\newcommand\fverb{\setbox\mybox=\hbox\bgroup\verb}
\newcommand\fverbdo{\egroup\medskip\noindent\fbox{\unhbox\mybox}\ }
\newcommand\fverbit{\egroup\item[\fbox{\unhbox\mybox}]}
\conference{Quantisations of exactly solvable ghostly models}
\abstract{We investigate an exactly solvable two-dimensional Lorentzian coupled quantum system that in a certain parameter regime can be transformed to a higher time derivative theory (HTDT) with preserved symplectic structure. By transforming the system's Lagrangian, we explicitly map it onto the standard Pais-Uhlenbeck formulation, revealing a direct correspondence in their dynamical and Poisson bracket structures. We quantise the model in two alternative ways. First we derive the eigensystem of the Hamiltonian by solving the Schr\"odinger equation through an Ansatz that leads to a set of coupled three-term recurrence relations, that we solve exactly, identifying normalisable wavefunctions and their associated energy spectra. We compare our results with a Fock space construction, finding exact agreement. On the basis of the exact solutions we report several specific physical properties of the ghost model investigated with a focus on the localisation properties of the system.}    

\title{Quantisations of exactly solvable ghostly models}
\author{Andreas Fring$^{\bullet}$, Takano Taira$^{\circ}$ and Bethan Turner$^{\bullet}$\\
	$\bullet$ Department of Mathematics, City St George's, University of London, \\ $\,\,$ Northampton Square, London EC1V 0HB, UK \\
	$\circ$ Department of Physics, Kyusyu University
	744 Motooka, Nishi-ku,  \\ $\,\,$ Fukuoka City, 819-0395, Japan\\ 
	
	e-mail: a.fring@city.ac.uk,taira.takano.292@m.kyushu-u.ac.jp,bethan.turner.2@city.ac.uk}

\input{tcilatex}
\begin{document}
	
\section{Introduction}	

Ghost-ridden quantum systems manifest in various forms. Typically, this means that parts of their spectra are not bounded from below or that they contain non-normalisable states, leading to a violation of unitarity. They may have linear momentum terms in their Hamiltonians, as for instance found in the dissipative Bateman oscillator \cite{bateman1931diss,dekker1981clas}  and Pais-Uhlenbeck oscillators \cite{pais1950field} in the Ostrogradsky scheme \cite{ostrogradsky1850memoire}, or exhibit non-positive definite kinetic energy terms of Lorentzian type \cite{diez2024foundations}. In this work, we investigate a model of the latter type.

In many cases the relation to the most prominent ghost ridden systems, higher time derivative theories (HTDT), is well known and often exploited in their analysis. These systems have undergone extensive investigations for more than half a century due their extremely attractive property of being renormalizable \cite{pais1950field,stelle77ren,grav1,grav2,grav3,modesto16super}. In particular, they have been considered as potential candidates for quantising gravity \cite{Hawking} and have also been proposed as models in various fields such as cosmology \cite{biswas2010towards,Salvio2,Salvio3,Salvio4}, finite temperature physics \cite{weldon98finite}, black hole physics \cite{mignemi1992black}, in a massless particle descriptions of bosons and fermions \cite{plyush89mass,Mpl} and in supersymmetric theories  \cite{dine1997comments,smilga17ultrav}. 

The formulation of consistent quantum versions of ghost ridden systems remains a challenge as they usually contain non-normalisable states and/or posses unbounded spectra. Several proposals have been made to cure these deficiencies \cite{ghostconst,salvio16quant,fakeons,bender2008no,raidal2017quantisation}. Various schemes to quantise finite particle theories have been explored, such as BRST quantisation \cite{rivelles2003triviality,Kap1,mandal2022bfv}, Dirac quantisation \cite{mann2005dirac,dama2006,bolonek2006comment,smilga2006ghost,smilga2009comments},
deformation quantisation \cite{berra2015def}, polymer quantisation \cite{cumsille2016polymer} or
imaginary scaling quantisation \cite{bender2008no,mosta2011im}. Schemes extending to field theories have also been explored, e.g. for scalar field theories \cite{Urries,weldon03quant,fring2024higher}.

In this work, we quantise a ghostly model in form of a two-dimensional Lorentzian coupled quantum system in two alternative ways. First, we directly solve the Schr\"odinger equation by making a suitable Ansatz that leads to a set of coupled three term recurrence relations. These relations are akin to the central set of equation occurring in quasi-exactly solvable models \cite{Bender:1995rh,Finkel,Krajewska,KhareM,E2Fring,AndTom5} and by appealing to techniques developed in that context we systematically solve them. Alternatively we use the fact that in a certain parameter regime the model can be symplectically mapped to the Pais-Uhlenbeck oscillator so that we can use its classical solutions as the basis for a Fock space construction. In the regime where the two systems coincide, we find perfect agreement between these two approaches.

Our manuscript is organised as follows: In section 2 we present the precise symplectic transformation from our ghost model to the Pais-Uhlenbeck oscillator. In section 3 and 4 we elaborate on the two alternative ways to quantise and solve the ghost system. In section 5 we study various physical properties of our model such as probability densities and uncertainty relations by comparing them also to the classical solutions. Our conclusions are presented in section 6.

\section{A ghostly model related to the Pais-Uhlenbeck oscillator}	
	
We consider a two-dimensional Hamiltonian system with Hamiltonian given by
\begin{equation}
	{\cal H}  =   \left(  p_x^2 - p_y^2  \right)  + \nu^2 x^2+  \Omega  y^2  + g x y, \qquad \nu,\Omega,g   \in \mathbb{R} . \label{Hghost}
\end{equation}
We note that the model is similar to a model recently proposed in  \cite{diez2024foundations} when the function in there is chosen to be quadratic. Since the kinetic term contains a negative term, we expect the model to be ``ghostly", in the sense that its spectrum might not be bounded from below or that unitarity is violated in its evolution. This is a feature shared with HTDTs, and in fact we argue that ${\cal H}$ can be viewed as a HTDT in disguise. 

To make this connection explicit, we first note that the Legendre transform 
\begin{equation}
	{\cal H}  =  \frac{\partial {\cal L} }{\partial \dot{x}} \dot{x}  + \frac{\partial {\cal L} }{\partial \dot{y}} \dot{y}  -{\cal L}.
\end{equation}
of the Lagrangian 
\begin{equation}
	{\cal L}  = \frac{1}{4} \left( \dot{x}^2  -\dot{y}^2  \right) -\nu ^2  x^2- \Omega  y^2 -g  x y,   \label{Lequ}
\end{equation}
leads to the Hamiltonian in (\ref{Hghost}). The classical dynamics for this model is described by the two coupled second order Euler-Lagrange equations resulting from (\ref{Lequ}) 
\begin{equation}
	\kappa_1 \left( \frac{1}{2}\ddot{x}+ g y + 2 \nu^2 x  \right)= 0, \qquad 
	\kappa_2  \left( \frac{1}{2} \ddot{y}- g x-  2 \Omega y \right) = 0. \label{equofm2}
\end{equation}
We have included here explicitly two constant overall factors $\kappa_1$ and  $\kappa_2$ for later purpose. In contrast, the Euler-Lagrange equations directly derived from the higher time-derivative PU Lagrangian in standard form 
\begin{equation}
	{\cal L}_{\text{PU}}  = \frac{1}{2}  \ddot{q}^2 -\frac{\zeta}{2}  \dot{q}^2+ \frac{\xi  }{2}  q^2, \quad \zeta, \xi \in \mathbb{R}^+   \label{LPU}
\end{equation}
is of fourth order
\begin{equation}
	   \ddddot{q} + \zeta \ddot{q} +\xi q =0 , \label{equofm}
\end{equation}
The standard parametrisation in terms of the frequencies $\omega_1$ and $\omega_2$ are $\zeta = \omega_1^2 +  \omega_2^2 $ and $\xi =  \omega_1^2  \omega_2^2$. There are multiple options to obtain (\ref{equofm}) from (\ref{equofm2}) and thus relate the corresponding dynamics in the respective phase spaces of the two models. Here, having the quantisation in mind, we choose an option that also preserves the symplectic structure, i.e. loosely speaking the Poisson bracket structure. Using the transformation map
\begin{equation}
 \Gamma: \quad	x \mapsto \mu_0 q+\mu_2 \ddot{q}, \qquad
	y \mapsto \nu_0 q+ \nu_2 \ddot{q},
	 \label{tttt}
\end{equation}
with constants
\begin{equation}
\mu_0 = \frac{\sqrt{2}  (g-2 \Omega )}{\sqrt{\nu ^2+\Omega-g}} , \qquad 
\nu_0 = \frac{\sqrt{2}  (g-2 \nu^2 )}{\sqrt{\nu ^2+\Omega-g}} , \qquad 
\mu_2 =-\nu_2 =\frac{1}{\sqrt{2} \sqrt{\nu ^2+\Omega -g}}, 
\end{equation}
\begin{equation}
	\zeta =   4\left( \nu ^2-\Omega \right), \qquad \xi =  4 \left(g^2-4 \nu ^2 \Omega \right) , \qquad
	\kappa_1 = \kappa_2= 2 \sqrt{2} \sqrt{\nu ^2+\Omega-g} .   \label{234}
\end{equation}
converts both equations (\ref{equofm2}) individually into the PU equation of motion (\ref{equofm}). Ensuring the reality of the coordinates restricts the parameter regime to $\nu ^2+\Omega>g$. At the same time this transformation preserves the Poisson bracket structure for the phase space variables $ \{x_1 =x, x_2 =y, p_1=p_x=\dot{x}/2,  p_2=p_y=-\dot{y}/2 \}$ compared to those for $\{ q, \dot{q}, \ddot{q} , \dddot{q}  \}$, see \cite{FFT} for details,
\begin{equation}
      \{  x_i , p_j    \}   = \delta_{ij}, \qquad     \Leftrightarrow \qquad  
      	\{ \dot{q}, \ddot{q}  \}_1=1, \quad   	\{ \dddot{q}, q  \}_1=1,  \quad   \{ \ddot{q}, \dddot{q}  \}_1= \zeta  .   \label{Poissxp}
\end{equation}
When directly transforming $	{\cal L}$ in (\ref{Lequ}) by means of (\ref{tttt}) we obtain $ 	{\cal L} =	{\cal L}_{\text{PU}} -\ddot{q}^2 - \dot{q} \ddddot{q} $. Thus, the two Lagrangians differ only by surface terms, which vanish upon integration, i.e. we may use $\int \ddot{q}^2  dt =- \int  \dot{q} \ddddot{q} dt $. 

It is well known that the PU oscillator exhibits a qualitatively different behaviour at its points of degeneracy, which is therefore a feature we also anticipate to find in our model. Comparing (\ref{equofm}) with the standard formulation of the PU equation in terms of the natural frequencies $\omega_1$ and $\omega_2$ allows to identify the degeneracy condition $\omega_1 =\omega_2 $ within the parameter space of our model (\ref{Hghost})
\begin{equation}
	\ddddot{x} + (\omega_1^2+ \omega_2^2) \ddot{x} +  \omega_1^2 \omega_2^2 x =0, \quad \Rightarrow \quad   \omega_1^2 = \omega_2^2, \quad  \Leftrightarrow \quad  \zeta^2 = 4 \xi    \quad  \Leftrightarrow \quad     g^2 =\left(\nu^2 + \Omega \right)^2  . \label{dege}
\end{equation}
We note that the point of degeneracy also marks the boundary of the validity for the transformation (\ref{tttt})-(\ref{234}) between our model in (\ref{Hghost}) and the PU-oscillator.
The point of degeneracy will play a critical role as a boundary value for the model domain in parameter space in shaping the spectral and dynamical properties of the system.

\section{Eigensystem from coupled three term recurrence relations}

We shall now solve the Schr\"odinger equation $H \psi_N(x,y)  = E_N \psi_N(x,y) $, $N=0,1,2, \ldots$ by making an Ansatz of Gaussian form dressed with polynomials in $x,y$ for the wavefunctions
\begin{equation}
	\psi_N(x,y) = P_N(x,y) e^{-\frac{\alpha  x^2}{2}-\frac{\beta  y^2}{2} +\gamma  x y   } , \qquad  P_N(x,y): =  
	 \sum _{i=0}^N    \sum _{j=0}^{ \lfloor \frac{N-i}{2} \rfloor }     \sigma_{i}^{N-2j}  x^i y^{N-i-2j} . \label{wansatz}
\end{equation}
Here ${\lfloor  z \rfloor}  := \max(n \in \mathbb{Z} \vert n \leq z  )  $ is the floor function that gives the greatest integer less than or equal to $z$. The Ansatz is similar to one previously used in \cite{felski2018analytic} to solve a coupled harmonic oscillator model. The procedure arising from this consists of solving coupled three term recurrence relations that arise from reading off the powers in $ x^i y^j $ upon the substitution of (\ref{wansatz}) into the Schr\"odinger equation.  

\subsection{The ``ground state"}
We start with the ``ground state", or more precisely the state with $N=0$, and an overall normalisation factor $P_0=\sigma_0^0$, which is in fact not always the state of lowest energy. A solution with energy $E_0 = \alpha - \beta$ is obtained when the model parameters $\nu, \Omega, g$ are related to the parameters entering via the wavefunction Ansatz, $\alpha, \beta, \gamma$, as   
\begin{equation}
	\alpha^2 - \gamma^2 = \nu^2, \qquad  \gamma^2 -\beta^2 = \Omega, \qquad
	g+2(\alpha - \beta) \gamma =0 ,  \label{abcg1}
\end{equation}
corresponding to the coefficients of $x^2 	\psi_0 $, $y^2 	\psi_0 $, and $x y	\psi_0 $, respectively.
\subsubsection{The non-degenerate case}

\begin{figure}[h]
	\centering         
	\begin{minipage}[b]{0.55\textwidth}           \includegraphics[width=\textwidth]{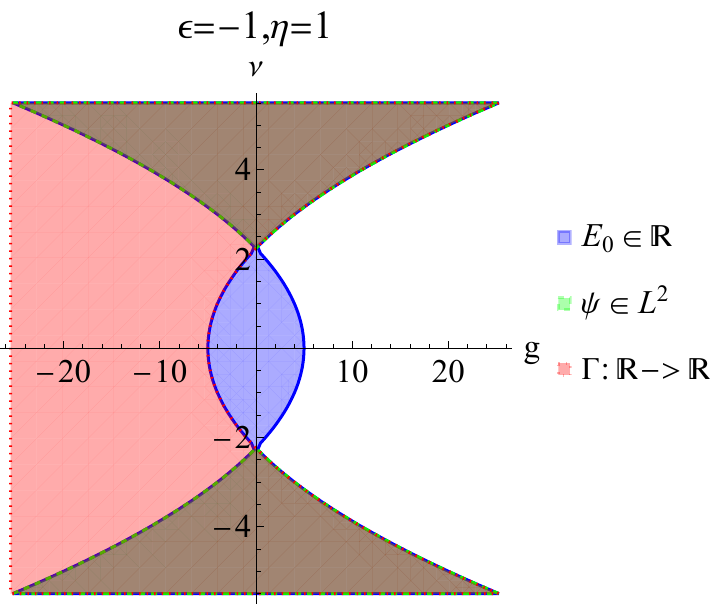}
	\end{minipage}   
	\begin{minipage}[b]{0.42\textwidth}           
		\includegraphics[width=\textwidth]{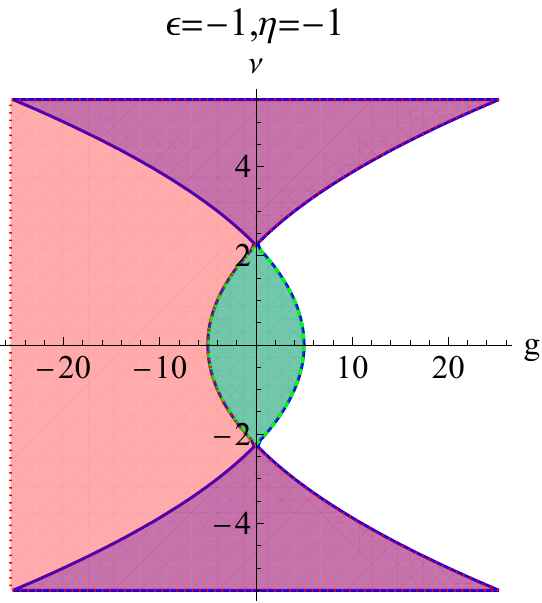}
	\end{minipage}    
	\begin{minipage}[b]{0.55\textwidth}           \includegraphics[width=\textwidth]{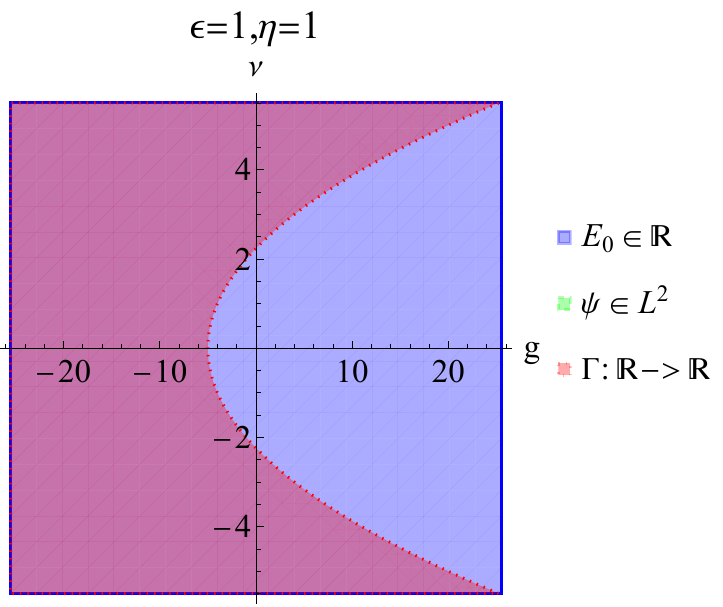}
	\end{minipage}   
	\begin{minipage}[b]{0.42\textwidth}           
		\includegraphics[width=\textwidth]{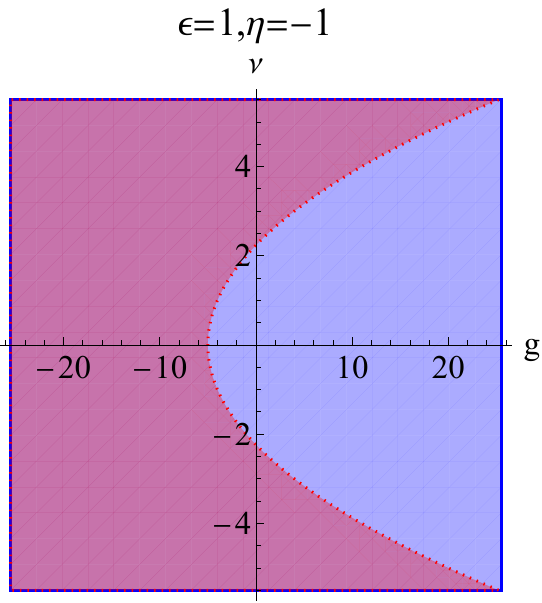}
	\end{minipage}    
	\caption{Parameter space domain for the three constraints real ground state energies, square integrability and real symplectic map to the PU oscillator, with $\Omega=-5$.}
	\label{PhysicalRegions}
\end{figure}

Since $\alpha, \beta, \gamma$ are just auxiliary variables we need to replace them by the actual model parameters $\nu, \Omega, g$. When solving the equations in (\ref{abcg1}) we obtain four different branches of the theory
\begin{equation}
	\alpha_\epsilon^\eta = \frac{2 \nu ^2  + \sigma_\epsilon }{ \Sigma_\epsilon^\eta}, \quad
	\beta_\epsilon^\eta = \frac{2 \Omega - \sigma_\epsilon}{\Sigma_\epsilon^\eta}, \quad
	\gamma_\epsilon^\eta = \frac{-g}{ \Sigma_\epsilon^\eta }, \,\, \Sigma_\epsilon^\eta = 2 \eta \sqrt{ \nu ^2-\Omega+ \sigma_\epsilon },
	\,\, \sigma_\epsilon =  \epsilon \sqrt{g^2-4 \nu ^2 \Omega } ,  \label{abcd}
\end{equation}
labelled by all combinations of $\epsilon = \pm 1$, $\eta=\pm 1$. We treat these solutions as different superselections sectors of the theory that are unrelated. Notice that for the degeneracy condition in (\ref{dege}) we have $\sigma_{-1} =-( \nu^2 - \Omega)$, so that $\Sigma_{-1}^\eta=0$ and the solutions in (\ref{abcd}) are not defined. This indicates the common feature that the degenerate case needs to be treated separately. Requiring further that $\psi_0(x,y) \in L^2(\mathbb{R}^2)$ is square integrable on $\mathbb{R}^2$ with $\alpha, \beta, \gamma \in \mathbb{R}$ leads to the constraints $\alpha > 0$, $\beta > 0$ and $\gamma^2 < \alpha \beta$\footnote{This is easily seen: The integral 
\begin{equation}
\int_{\mathbb{R}^2 } \left\vert  e^{-\frac{\alpha  x^2}{2}-\frac{\beta  y^2}{2} +\gamma  x y   }  \right\vert^2 dx dy=
\int_{\mathbb{R}^2 }  e^{-(x,y) M (x,y)^{ \intercal}} dx dy, \qquad M= \left(
\begin{array}{cc}
	\alpha  & -\gamma  \\
	-\gamma  & \beta  \\
\end{array}
\right)
\end{equation}
is finite when $M$ is positive definite, i.e. when $\alpha > 0$, $\beta > 0$ and $\det M >0 \Leftrightarrow\gamma^2 < \alpha \beta$. }.
This allows us to exclude the two cases with $\epsilon =1$, since then the inequalities can not be satisfied in the parameter space, leaving the two options with $\epsilon =-1$.  In addition, we demand the energy $E_0$ to be real so that $\Sigma_\epsilon^\eta, \sigma_\epsilon  \in \mathbb{R} $. Combining these requirement restricts the parameter space domain in the two remaining cases to
\begin{eqnarray}
 \epsilon=-1, \eta=-1 &:& \quad \vert   g \vert  <- \nu^2 - \Omega  ,  \label{reconst} \\ 
  \epsilon=-1, \eta=1 &:&  \quad \vert   g \vert  < \nu^2 + \Omega \,\,  \land \,\,  \vert   g \vert  < 2 \nu^2 \,\,  \land \,\,  g^2 > 4 \nu^2 \Omega .
\end{eqnarray}
In both of these cases we can find regions in parameter space that satisfy these inequalities as depicted in figure \ref{PhysicalRegions}.  

We notice that in both cases the strong coupling regime, i.e. large $g$, does not admit normalisable solutions with real ``ground state" energy. Furthermore, we observe that when enforcing the map $\Gamma$ in (\ref{tttt}) to be real and well-defined, only leaves the brown region for the case $\epsilon = - \eta =1$, whereas for the case $\epsilon = \eta =-1$ we have normalisabilty and real ground state energies, but can not make contact to the PU-oscillator in any region of the parameter space in a well-defined manner. For both cases with $\epsilon=1$ the reality of the ground state energy together with the validity of the map $\Gamma$ can be guaranteed in large parts of the parameter regime, but the wavefunctions are never normalisable. We summarize these features in table \ref{table1}, indicating by a $\checkmark$-sign that regions in parameter space can be found in which the three conditions on the left hold and a $\times$-sign when the conditions lead to an empty set in parameter space.

	\begin{table}[h]
\begin{tabular}{|c||c|c|c|c|}
	\hline
  	& $\epsilon=-\eta =-1$ & $\epsilon=\eta =-1$ & $\epsilon=\eta =1$ & $\epsilon=-\eta =1$ \\
	\hline \hline
$E_0 \in \mathbb{R}$, $\psi_0 \in L^2$, $\Gamma: \mathbb{R} \mapsto	 \mathbb{R}$& \checkmark & $\times$ & $\times$ & $\times$  \\
	\hline
$E_0 \in \mathbb{R}$, $\psi_0 \in L^2$ & \checkmark & \checkmark & $\times$ & $\times$  \\
	\hline
$\psi_0 \in L^2$, $\Gamma: \mathbb{R} \mapsto	 \mathbb{R}$	& \checkmark  & \checkmark  & $\times$ & $\times$ \\
	\hline
$E_0 \in \mathbb{R}$, $\Gamma: \mathbb{R} \mapsto	 \mathbb{R}$	  & \checkmark & \checkmark & \checkmark & \checkmark \\
	\hline
\end{tabular}
\caption{Non-empty ($\checkmark$) and empty ($\times$) domains in parameter space when imposing combinations of the constraints related to the reality of the ground state energy, square integrability or a well-defined map from the ghost model to the PU-oscillator.} \label{table1}
\end{table}

It is clear from the Ansatz for $\psi_N$, that the constraints resulting from the normalisability requirement also hold for $N>0$, as in that case the wavefunction only differs by an overall polynomial factor that does not affect the convergence behaviour. Below we see that also the constraints (\ref{abcg1}) remain valid, so that (\ref{reconst}) holds for all wavefunctions, but additional constraint may arise to ensure the reality of the spectra.

\subsubsection{The degenerate case}

Solving (\ref{abcg1}) for $\alpha, \beta, \gamma$ in the frequency degenerate case when $g = \eta (\nu^2 + \Omega ) $ yields the four solutions
\begin{equation}
	\alpha_\epsilon= \epsilon \frac{  \Omega -3 \nu ^2}{2 \sqrt{2} \sqrt{\nu ^2-\Omega }}, \quad
	\beta_\epsilon =\epsilon  \frac{  \nu ^2-3 \Omega }{2 \sqrt{2} \sqrt{\nu ^2-\Omega }}, \quad
	\gamma_\epsilon^\eta = \eta  \epsilon \frac{ \nu ^2+\Omega }{2 \sqrt{2} \sqrt{\nu ^2-\Omega }}  ,  \label{abcdeg}
\end{equation}
In this case the constraint $\alpha > 0$, $\beta > 0$ and $\gamma^2 < \alpha \beta$ can not be satisfied for \emph{any} of the four solutions, which in turn implies that the wavefunctions can not be normalised in the degenerate case. The reality of the energy $E_0$ is guaranteed in this case with the simple constraint $\nu ^2 > \Omega$. The map $\Gamma$ is also not defined in this case.

 \subsection{The three term recurrence relations} 

Upon the validity of (\ref{abcg1}) and fixing $N$, we obtain a three term recurrence relation in $n$ for the coefficients $\sigma_n^{N}$  and $\lfloor N/2 \rfloor$ three term recurrence relations with a non-homogenous term for the coefficients $\sigma_n^{N-2k}$ with $k=1,\ldots$. The relations read
\begin{equation}
	a_n^{N,k}  \sigma_{n-1}^{N-2k}   +  b_n^{N,k} \sigma_n^{N-2k}  +  c_n \sigma_{n+1}^{N-2k} 
 = f_n^{N,k} \sigma_n^{N-2 k+2}+ g_n \sigma_{n+2}^{N-2 k+2},   \label{recac} 
\end{equation}
where the coefficients are
\begin{eqnarray}
	a_n^{N,k} &:=&  2 (2k+n-N-1)  \gamma, \quad 	b_n^{N,k} := E - (2n+1) \alpha + \left(2N-4k -2 n +1 \right) \beta, \\
	c_n &:=&  2 (n+1) \gamma, \,\,  f_n^{N,k} := (N-2 k-n+1) (N-2 k-n+2), \,\, g_n:= -(n+1) (n+2), \notag
\end{eqnarray}	
with $\sigma_n^m = 0$ for $n=0, \ldots, N$, $k=0, \ldots, \lfloor (N-n)/2 \rfloor$ and  $N<n< 0$, $ \lfloor (N-n)/2 \rfloor<m <0  $.  We stress that when reading off powers in $x$ and $y$ there will be additional terms besides those reported in (\ref{recac}), which, however, all cancel out when imposing the constraints (\ref{abcg1}). 

Defining the $(N+1)$-dimensional vector $\Sigma_N^k = (\sigma_0^{N-2k},\sigma_1^{N-2k},\ldots,\sigma_{N-1}^{N-2k},\sigma_N^{N-2k})$ and the matrices 
\begin{equation}
	M_N^k= \left(
	\begin{array}{cccccc}
		b_0^{N,k}  & c_0 & 0  & \cdots& 0 & 0 \\
	a_1^{N,k} & b_1^{N,k}   &c_1 &  \cdots & 0 & 0 \\
		0 & a_2^{N,k}&b_2^{N,k}    &\ddots & 0 & \vdots \\
		0 & 0 &  \ddots& \ddots  &  \ddots & \vdots \\
		\vdots & \vdots & \ddots & \ddots&  0& 0 \\
		0 & 0 & \ddots  &  a_{N-1}^{N,k}   & b_{N-1}^{N,k}   & c_{N-1} \\
		0 & 0 &\cdots  & 0 & a_N^{N,k} & b_N^{N,k}  \\
	\end{array}
	\right), \,
		B_N^k= \left(
	\begin{array}{ccccccc}
		f_0^{N,k}  & 0& g_0 & 0 & \cdots& 0 & 0 \\
	0& f_1^{N,k}   &0 & g_1 &  \cdots & 0 & 0 \\
		0 & 0&\ddots  &\ddots  &\ddots& 0 & \vdots \\
		0 & 0 &  \ddots& \ddots &  \ddots &  \ddots &0 \\
		\vdots & \vdots & \ddots &  \ddots& \ddots&  0& g_{N-2} \\
		0 & 0 & \ddots & 0 &  0   & f_{N-1}^{N,k}   & 0 \\
		0 & 0 &\cdots & \cdots & 0 & 0 & f_N^{N,k}  \\
	\end{array}
	\right),
\end{equation}
it is convenient to cast the recurrence relations (\ref{recac}) into matrix form
\begin{equation}
	   M_N^k  \left( \Sigma_N^k \right)^\intercal   =  B_N^k  \left( \Sigma_N^{k-1} \right)^\intercal , \qquad  k=1, \ldots , \left\lfloor \frac{N-i}{2}  \right\rfloor  .
\end{equation}
These equations are easily solved to
\begin{equation}
  \left( \Sigma_N^k \right)^\intercal   =  \prod_{\ell=1}^{k}  \left[   \left( 	M_N^\ell       \right)^{-1}            B_N^\ell \right] \left( \Sigma_N^0 \right)^\intercal  .  \label{closedsol}
\end{equation}
Next we discuss how to find the initial set of coefficients $\Sigma_N^0 $ and elaborate on how to compute the inverse of $M_N^\ell  $ efficiently by exploiting the fact that they are tridiagonal matrices.

\subsubsection{Energy quantisation from the $k=0$ recurrence relation}
\begin{figure}[h]
	\centering         
	\begin{minipage}[b]{0.56\textwidth}           \includegraphics[width=\textwidth]{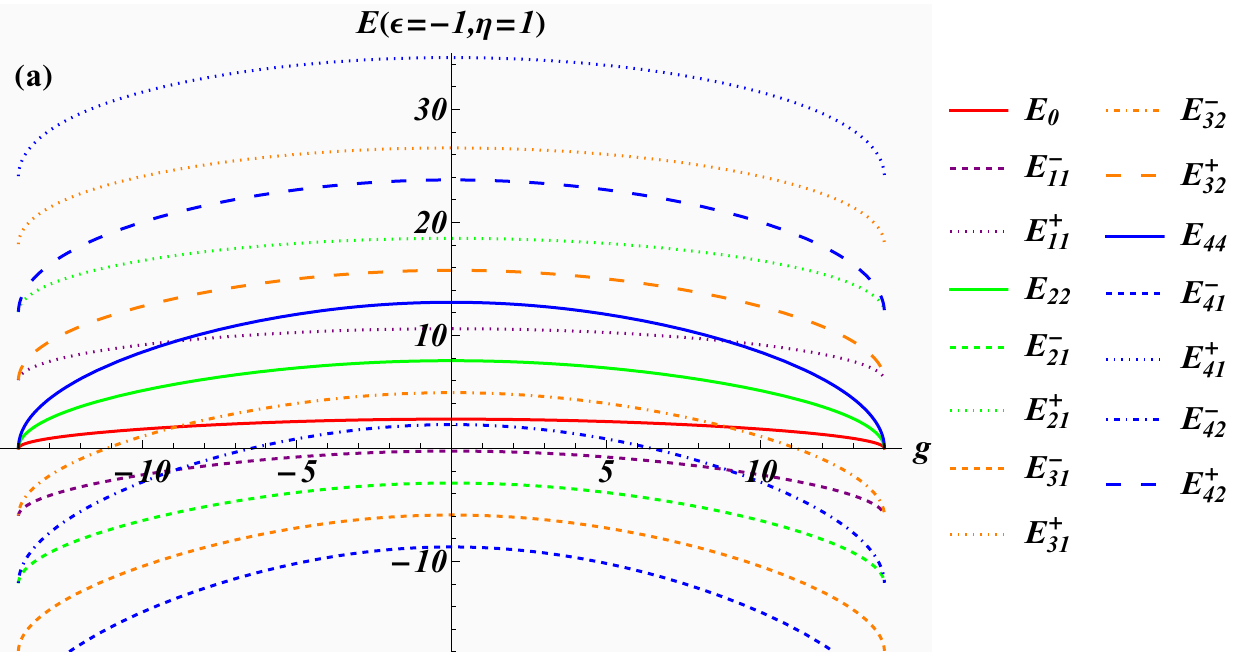}
	\end{minipage}   
	\begin{minipage}[b]{0.42\textwidth}           
		\includegraphics[width=\textwidth]{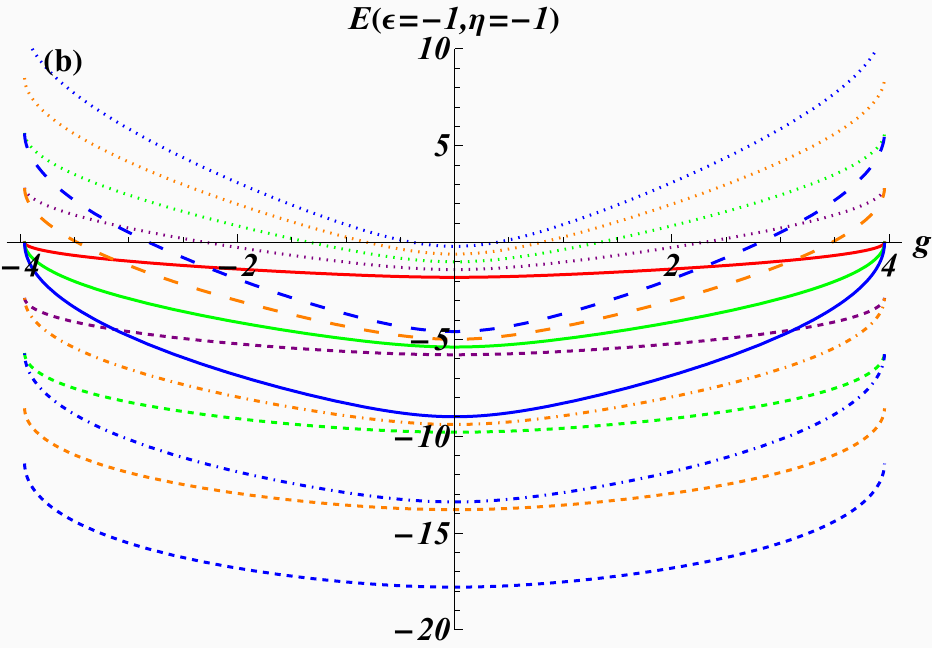}
	\end{minipage}    
	\begin{minipage}[b]{0.56\textwidth}           \includegraphics[width=\textwidth]{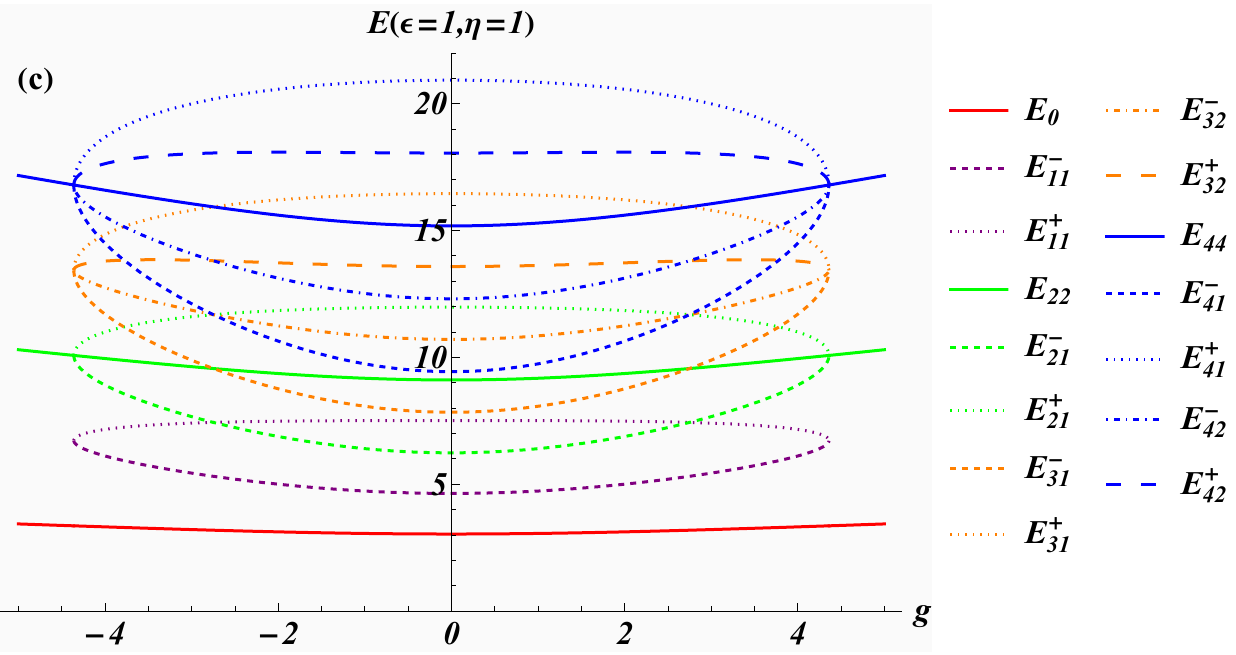}
	\end{minipage}   
	\begin{minipage}[b]{0.42\textwidth}           
		\includegraphics[width=\textwidth]{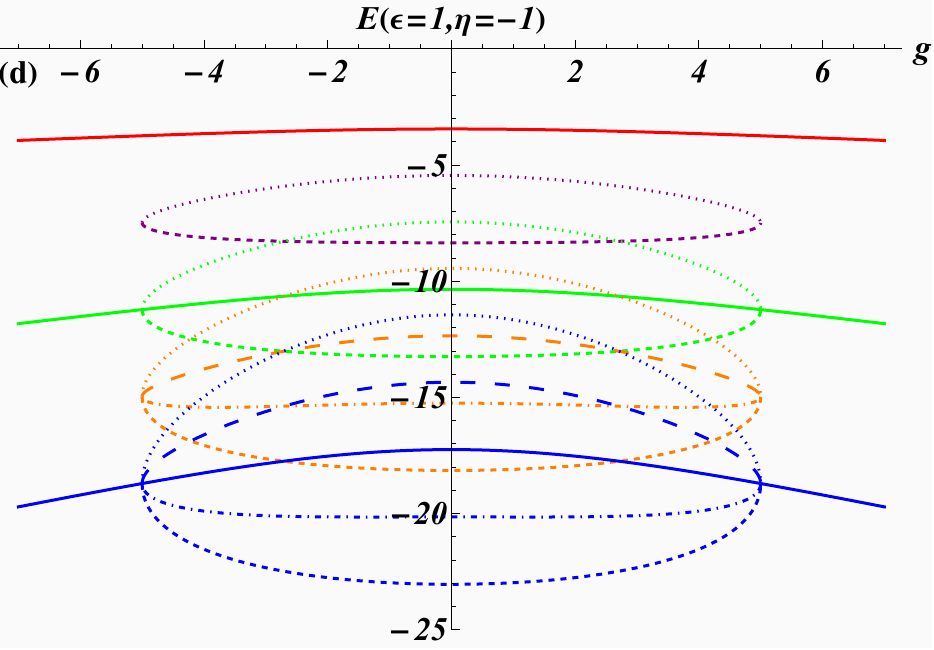}
	\end{minipage}    
	\caption{Real energy spectra for normalisable eigenstates in panels (a) and (b) with $\nu=4$, $\Omega = -2$, $\vert g \vert < 14 $ and $\nu=0.2$, $\Omega = -4$, $\vert g \vert < 3.96 $ and for non-normalisable eigenstates in panels (c) and (d) with $\nu=0.8$, $\Omega = -5$ and $\nu=1$, $\Omega = -6$, respectively. }
	\label{EnSpec}
\end{figure}

In the first relation of (\ref{recac}) for $k=0$ the term on the right hand side is vanishing so that the relation reduces to a genuine three term relation in $n$
\begin{equation}
-2 (N+1-n) \gamma \sigma_{n-1}^{N}   + [E - (2n+1) \alpha + (2N -2 n +1) \beta]   \sigma_n^{N}  + 2 (n+1) \gamma \sigma_{n+1}^{N}  =0. \label{recthreet}
\end{equation}
These relations are the key characteristic for quasi-exactly solvable models \cite{Bender:1995rh,Finkel,Krajewska,KhareM,E2Fring,AndTom5}, allowing us to exploit that structure to solve for all coefficients $\sigma_n^N$. Typically the three-term relation reduces to a two-term relation at certain values of $n$, in our case when $n=N+1$, which enforces a quantisation condition for the energy $E$ on which we elaborate at first.
For the matrix relation $M_N^0  \left( \Sigma_N^0 \right)^\intercal =0$ to have non-trivial solutions we require that the determinant $\vert M_N^0 \vert$ is vanishing. Noting that $M_N^0$ is a tridiagonal matrix, we can compute its determinant recursively from
\begin{equation}
      R^{N,0}_{t+1} = b_{t-1}^{N,0} R^{N,0}_{t} -  a_{t-1}^{N,0} c_{t-2}^{N,0} R^{N,0}_{t-1},
\end{equation}
with $R^{N,0}_{n \leq 0}=0$, $R^{N,0}_1=1$ for $t=1, \ldots , N+1$  where $R^N_t$ denotes the determinant of the $ (t-1) \times (t-1)$ leading principal submatrix in the top left corner. Hence $ \vert  M_N^0   \vert = R^N_{N+2}$, see e.g. \cite{usmani94inv}. Solving this equation we obtain the conveniently factorised expression for the determinant 
\begin{equation}
\vert  M_N^0   \vert =  g\left(E,N,\frac{N}{2} +1 \right)^{\frac{(N+1 \bmod 2)}{2}} \prod _{k=n}^{\left\lfloor \frac{N+1}{2}\right\rfloor } g(E,N,n),
\end{equation}
where the function $g(E,n,m)$ is defined as
\begin{equation}
   g(E,n,m) =  [E+  (2 m-2 n-3)\alpha +  (2 m-1) \beta] [E  +(1-2 m) \alpha +  (3-2 m+2 n) \beta]
   +4 \gamma ^2 (2-2m+n)^2,
\end{equation}
with $(N+1 \bmod 2)$ equal $1$ or $ 0$ for N even or odd, respectively and $\lfloor x \rfloor$ denoting the floor function defined above.  Thus the determinant of $M_N^0$ is vanishing whenever the quadratic equations is solved in $E$, i.e. $ g(E,n,m) = 0$, namely for the energies
\begin{eqnarray}
	E_{Nn}^\pm &=& (N+1) (\alpha -\beta )\pm (2-2 n+N)\sqrt{ (\alpha +\beta )^2-4 \gamma ^2}, \quad n=1, \ldots , \left\lfloor \frac{N}{2}\right\rfloor , \label{eab1} \\
	E_{NN} &=&  (1+N) (\alpha -\beta ), \qquad     (N+1 \bmod 2). \label{eab2}
\end{eqnarray}
We depict a few sample spectra in figure \ref{EnSpec}, where we distinguish between the four branches. First of all we notice that the spectra for the two normalisable solutions are not bounded from below, whereas one of those corresponding to the non-normalisable cases is bounded from below and the other from above.

 Having fixed all parameters of the model but one, i.e. the coupling strength $g$ between the two oscillators, we observe the typical avoided level crossing \cite{vNW} for fixed values of $N$. However, the level may cross for different values of $N$, which indicates that they do not couple because they belong to different symmetry classes. These will be identified below. 
 
At the boundary of the domain the normalisable and non-normalisable solutions exhibit qualitatively distinct behaviour as is seen from the limit to the boundary value 
\begin{eqnarray}
	\lim_{g \rightarrow \nu^2 +\Omega} E_{N,n}^{\pm} &=& (N+1 ) \eta \epsilon 
	\sqrt{\nu^2 -\Omega} + \epsilon \vert \nu^2 -\Omega    \vert  \pm (1-\epsilon)
	\left(  1-n +\frac{N}{2}      \right)     \sqrt{2 (\nu^2  - \Omega)},  \qquad \,\, \\
	\lim_{g \rightarrow \nu^2 +\Omega} E_{N,N}^{\pm} &=& (N+1 ) \eta \epsilon 
	\sqrt{\nu^2 -\Omega} + \epsilon \vert \nu^2 -\Omega    \vert .
\end{eqnarray}
We observe that for $\epsilon=-1$ the values for the energies $E_{N,n}^{+}$ and $E_{N,n}^{-}$ are distinct, whereas for $\epsilon=-1$ the spectra develop an exceptional point  \cite{Kato,berry2004physics,miri2019exceptional} at the boundary with both values becoming identical to $E_{N,N}$ and complex conjugate pairs thereafter. This is the well known behaviour for $\cal{PT}$-symmetric quantum mechanical systems where the transition from the $\cal{PT}$-symmetric to the spontaneously broken regime occurs at the exceptional point \cite{PTbook}, albeit here for a real Hermitian Hamiltonian. The Hamiltonian is evidently $\cal{PT}$-symmetric under $\cal{PT}:$ $p_x \rightarrow p_x$, $p_y \rightarrow p_y$, $x \rightarrow -x$ and $y \rightarrow -y$ and the condition that the eigenstates of $\cal{H}$ are also eigenstates of the $\cal{PT}$-operator no longer holds in the broken regime when the parameters $\alpha, \beta, \gamma$ become complex.

Notice that in the frequency degenerate case (\ref{abcdeg}) we always have $(\alpha +\beta )^2-4 \gamma ^2=0$ throughout the entire parameter regime, so that also the energies are indeed degenerate.

\subsubsection{Solution of the $k=0$ three term recurrence relation}

Next we provide a closed solution for the coefficients $\sigma_n^N$ following \cite{Gonoskov,E2Fring} in solving the recurrence relations. For this purpose we first transform (\ref{recac}) to the canonical form for three-term recurrence relation 
\begin{equation}
	\tau_{n+1} = \tau_n + \lambda_n \tau_{n-1}  \label{taurec},
\end{equation}
with
\begin{equation}
	\sigma_n^N:= \kappa \tau_{n}\prod_{k=0}^{n-1}  r_k  =   \kappa \tau_{n} \frac{ \left[ \alpha -(2N +1) \beta -E \right]  }{2 n! (\alpha +\beta )  }    
		\frac{\Gamma \left[ n-\frac{E-\alpha +(2 N +1) \beta }{2 (\alpha +\beta
				)}\right]}{\Gamma \left[ \frac{3 \alpha -(2 N-1) \beta -E }{2 (\alpha +\beta )}\right] } \left(\frac{\alpha +\beta }{\gamma }\right)^n
\end{equation}
where $\kappa$ is an overall constant and 
\begin{equation}
 r_n:=- \frac{b_n^{N,0}}{c_n^{N,0}}  \qquad
 s_n:= - \frac{a_n^{N,0}}{c_n^{N,0}} , \qquad 
 \lambda_n :=  \frac{s_n}{ r_{n-1} r_n  }.
\end{equation}
The recurrence relation (\ref{taurec}) is solved by
\begin{equation}
	\tau_n = 1+ \sum_{p=1}^{\lfloor  n/2 \rfloor} S(n-1, p),
\end{equation}	
where ${\lfloor  x\rfloor} $ is defined as above and $S(n-1, p)$ as the $p$-fold sum
\begin{equation}
	S(n,p) := \sum_{k_1=1}^{n+2(1-p)} \lambda_{k_1} \sum_{k_2=2+k_1}^{n+2(2-p)} \lambda_{k_2} \cdots \sum_{k_{p-1}=2+k_{p-2}}^{n-2} \lambda_{k_{p-1}} \sum_{k_p=2+k_{p-1}}^{n} \lambda_{k_p}.	
\end{equation}	
The first expressions are computed to $ \tau_0 = 1 $, $ \tau_1 = 1$, $\tau_2 = 1 + \lambda_1 $, 
$ \tau_3 = 1 + \lambda_1 + \lambda_2$, 
$\tau_4 = 1 + \lambda_1 + \lambda_2 + \lambda_3 + \lambda_1 \lambda_3 $, 
$\tau_5 = 1 + \lambda_1 + \lambda_2 + \lambda_3 + \lambda_4 + \lambda_1 \lambda_3 + \lambda_1 \lambda_4 + \lambda_2 \lambda_4 $ etc. Thus we compute the first coefficients to
\begin{eqnarray}
	\sigma_0^N &=& \kappa,   \\ 
	\sigma_1^N &=& \kappa   r_0,  \notag \\ 
	\sigma_2^N &=& \kappa \left( s_1 +   r_0  r_1      \right) ,\notag  \\ 
	\sigma_3^N &=& \kappa  \left( s_1 r_2+   s_2  r_0 + r_0 r_1 r_2      \right), \notag \\ 
	\sigma_4^N &=& \kappa \left( s_1 s_3 +   s_2  r_0 r_3+ s_1 r_2 r_3  + r_0 r_1 r_3 + r_0 r_1 r_2   r_3      \right), \notag    \\ 
    \sigma_5^N &=& \kappa \left( s_1 s_3 r_4 + s_1 s_4 r_2 + s_2 s_4 r_0 + s_1 r_2 r_3 r_4 + s_2 r_0 r_3 r_4 
    + s_3 r_0 r_1 r_4 + s_4 r_0 r_1 r_2 +  r_0 r_1 r_2 r_3 r_4       \right).   	 \notag
\end{eqnarray}	
We have now fully determined the vector $\Sigma_N^0$.

\subsubsection{Solutions of the non-homogeneous three term recurrence relations, $k \neq 0$}

Closed formal solutions to the non-homogeneous three term recurrence relations (\ref{recac}) with $k=1,2, \ldots$ were found in (\ref{closedsol}). The expressions still involve the inverse of the matrix $M_N^\ell$, which can be computed efficiently by exploiting the fact that it is a tridiagonal matrix admitting the closed formula 
\begin{equation}
\left(M_N^\ell \right)^{-1}_{ij} =
\begin{cases}
	\displaystyle (-1)^{i+j} \frac{R_{i}^{N,\ell} T_{j+1}^{N,\ell}}{\vert  M_N^0   \vert   } \prod_{k=i}^{j-1} c_{k-1}^{N,\ell}, & \text{if } i < j, \\[12pt]
	\displaystyle \frac{R_{i}^{N,\ell} T_{i+1}^{N,\ell}}{\vert  M_N^0   \vert }, & \text{if } i = j, \\[12pt]
	\displaystyle (-1)^{i+j} \frac{R_{j}^{N,\ell} T_{i+1}^{N,\ell}}{\vert  M_N^0   \vert } \prod_{k=j}^{i-1} a_k^{N,\ell}, & \text{if } i > j.
\end{cases}
\end{equation}
Here the $R_{t}^{N,\ell}$ are the determinants of the $ (t-1) \times (t-1)$ leading principal submatrix in the top left corner of $M_N^\ell$ computed recursively from
\begin{equation}
	R^{N,\ell}_{t+1} = b_{t-1}^{N,\ell} R^{N,\ell}_{t} -  a_{t-1}^{N,\ell} c_{t-2}^{N,\ell} R^{N,\ell}_{t-1},
\end{equation}
with $R^{N,\ell}_{n \leq 0}=0$, $R^{N,\ell}_1=1$ for $t=1, \ldots , N+1$. The $T_{t}^{N,\ell}$ are determinants of the $ (t-1) \times (t-1)$ trailing principal submatrix in the lower right corner of $M_N^\ell$ starting at row $t$, which can be computed from
\begin{equation}
T^{N,\ell}_{t} = b_{t-1}^{N,\ell} T^{N,\ell}_{t+1} -  a_{t}^{N,\ell} c_{t-1}^{N,\ell} T^{N,\ell}_{t+2},
\end{equation}
with $T^{N,\ell}_{n >  N+2}=0$, $T^{N,\ell}_{N+2}=1$ for $t=N, N-1 \ldots , 1$.

Note that the quantisation condition has selected the energy $E$ such that $\vert  M_N^0   \vert  =0$, which in turn ensures that  $\vert  M_N^\ell   \vert  \neq 0$ for $\ell=1,2,\dots$ so that all relevant inverse matrices in the procedure exist.  

\subsection{First explicit examples}

Having found the entire eigensystem it is useful to report some explicit examples. The first solutions for the polynomials $P_N$ are easily found. Up to an overall multiplicative constant $c_0^N$ they read 
\begin{eqnarray}
P_0 &=& 1,\\
P_1  &=&    y-\frac{ E-\alpha +3 \beta }{2 \gamma } x , \notag  \\
P_2  &=&  x^2+y^2+\frac{(E-3 \alpha +3 \beta ) (E-\alpha +5 \beta ) \left[x^2 (E-\alpha +\beta )-2\right]}{8 \gamma ^2 (E-\alpha +\beta )}-\frac{ (E-\alpha +5 \beta )}{2 \gamma }  xy .\notag \\
\end{eqnarray}
Together with the energies from the quantisation condition (\ref{eab1}) and (\ref{eab2}), the lowest level eigensolutions are
\begin{eqnarray}
	N=0 &:&  P_0 =1 , \quad E_0 = \alpha -\beta ,\\
	N=1 &:&  P_1^1  =y-\frac{ \alpha +\beta -\kappa }{2 \gamma } x, \quad E_1^1 = 2 ( \alpha -\beta) -\kappa ,\\
	         &&   P_1^2  =y-\frac{\alpha +\beta +\kappa }{2 \gamma } x , \quad E_1^2 = 2 ( \alpha -\beta) +\kappa , \notag \\
	N=2 &:&  P_2^1  =  x^2 +y^2 -\frac{\alpha +\beta }{\gamma } xy , \quad E_2^1 = 3 ( \alpha -\beta), \\    
	        &&  P_2^2  =  \frac{\kappa^2 + (\alpha + \beta) \kappa}{2 \gamma ^2 (\beta-\alpha  +\kappa )}+x^2 \left[  \frac{(\alpha +\beta ) (\alpha +\beta -\kappa )}{2 \gamma
	        	^2}-1\right]-\frac{(\alpha +\beta -\kappa )}{\gamma } x y +y^2,  \notag  \\ 
	         &&  P_2^3 =\frac{\kappa^2 - (\alpha + \beta) \kappa}{2 \gamma ^2 (\beta-\alpha  -\kappa )}+x^2 \left[  \frac{(\alpha +\beta ) (\alpha +\beta +\kappa )}{2 \gamma
	         	^2}-1\right]-\frac{(\alpha +\beta +\kappa )}{\gamma } x y +y^2,  \qquad   \notag \\ 
	         &&    E_2^2 = 3 ( \alpha -\beta) -2 \kappa, \qquad  E_2^3= 3 ( \alpha -\beta)+ 2 \kappa,  \notag 
\end{eqnarray}
with $\kappa:= \sqrt{(\alpha +\beta )^2-4 \gamma ^2} $.

\section{Bosonic Fock space construction}
Next we second quantise the model and compare with the results obtained in the previous section. Our starting point is the well-known mode expansion solution for the PU-equation \cite{pais1950field} in the non-degenerate case
\begin{equation}
	q(t) =a e^{-i t \omega _1}+ a^\star  e^{i t \omega _1} + b e^{-i t \omega _2}+ b^\star  e^{i t \omega _2} , \qquad 
	a,a^\star ,b, b^\star \in \mathbb{C}.  \label{regsol}
\end{equation}
Imposing the frequencies to be real implies that the parametrisation (\ref{dege}) limits the regimes to $0<\zeta$, $0<\xi< \zeta^2/4$. In model parameter space these values correspond to the restrictions $\nu^2 > \Omega$, $g^2 > 4 \nu^2 \Omega$, $g^2 > (\nu^2 + \Omega)^2  $, which we already encountered previously. Expressing the frequencies in terms of the model parameters gives rise to eight different regimes depending on their respective ordering and signs. We obtain
\begin{equation}
	 \omega_i^{\epsilon \eta} =\frac{\epsilon}{ \sqrt{2} }   \sqrt{ \zeta + \eta \sqrt{\zeta^2 - 4 \xi}         },
\end{equation}
with $\zeta$ and $\xi$ unambiguously defined in (\ref{234}), so that we have the regimes 
{\small 
\begin{align}
	\text{I} :    \omega_2^{++} > \omega_1^{+-} >0,      &  \,\,\,\,	\text{II}:    \omega_1^{++} > \omega_2^{+-}>0,   & \text{III}:   \omega_2^{++} > - \omega_1^{--} >0,     & \,\,\,\, \text{IV}:  - \omega_1^{-+} > \omega_2^{+-}>0, \notag  \\
		\text{V} :   0>  \omega_2^{--} > \omega_1^{-+},   &	 \,\, \text{VI}:     0>  \omega_1^{--} > \omega_2^{-+},  & \text{VII}:  0>  \omega_2^{--} >- \omega_1^{++},    & \,\, \text{VIII}:  0> - \omega_1^{+-} > \omega_2^{-+}. \notag 
\end{align}  }
 Using the solution (\ref{regsol}) the transformation (\ref{tttt}) then yields by direct calculation
\begin{eqnarray}
	x &=&  \bar{\mu}_1 \left(  a e^{-i t \omega _1}+ a^\star  e^{i t \omega _1}   \right)  +\bar{\mu}_2    \left(   b e^{-i t \omega _2}+ b^\star  e^{i t \omega _2}    \right) , \label{clsol1} \\
	y &=&    \bar{\nu}_1 \left(  a e^{-i t \omega _1}+ a^\star  e^{i t \omega _1}   \right)  +  \bar{\nu}_2    \left(   b e^{-i t \omega _2}+ b^\star  e^{i t \omega _2}    \right) , \\
	p_x   &=& \frac{i}{2} \omega _1 \bar{\mu}_1 \left(a^* e^{i t \omega _1}-a e^{-i t \omega _1}\right)+\frac{i}{2} \omega _2 \bar{\mu}_2 \left(b^* e^{i t \omega _2}-b e^{-i t \omega _2}\right) , \quad \\
	p_y  &=&  \frac{i}{2} \omega _1 \bar{\nu}_1\left(a e^{-i t \omega _1}-a^* e^{i t \omega _1}\right)+\frac{i}{2}  \omega _2 \bar{\nu}_2  \left(b e^{-i t \omega _2}-b^* e^{i t \omega _2}\right),  \label{clsol4}
\end{eqnarray}
with $\bar{\mu}_i := \mu _0- \mu _2 \omega _i^2 $,  $\bar{\nu}_i := \nu _0- \nu _2 \omega _i^2 $. When solved for $a,a^\star ,b, b^\star $ this gives
\begin{eqnarray}
	a &=&   \left[  \omega _1 \bar{\mu }_2 y  - \omega _1 \bar{\nu }_2 x    -2 i  ( \bar{\nu }_2 p_x +  \bar{\mu }_2   p_y  )  \right]
	\frac{e^{i t \omega _1}}{2 \omega _1  \left( \omega_1^2 -  \omega_2^2   \right)      \left(\mu _2 \nu_0-  \mu_0 \nu_2 \right) } , \label{comm1}  \\ 	
	b &=&    \left[  \omega _2 \bar{\nu }_1 x   -\omega _2 \bar{\mu }_1 y   +2 i  ( \bar{\nu }_1 p_x +  \bar{\mu }_1   p_y  )  \right]
	\frac{e^{i t \omega _2}}{2 \omega _2  \left( \omega_1^2 -  \omega_2^2   \right)      \left(\mu _2 \nu_0-  \mu_0 \nu_2 \right) }  ,   \label{comm2}  
\end{eqnarray}
and $a^\star ,b^\star $ simply being the complex conjugates of $a,b$. Next we promote $x,y,p_x,p_y$ and $a,a^\star ,b, b^\star $  to operators  $\hat{x},\hat{y},\hat{p}_x,\hat{p}_y$ and $\hat{a},\hat{a}^\dagger ,\hat{b},\hat{b}^\dagger  $, respectively. The Poisson brackets (\ref{Poissxp}) become equal time commutators $ [ \hat{x}_j , \hat{p}_k   ]  = i  \delta_{jk}$ provided that 
\begin{eqnarray}
     \left[ \hat{a}, \hat{a}^\dagger  \right]   &=&   \frac{\nu_0^2- \mu_0^2 + 2( \mu_0 \mu_2 -\nu_0 \nu_2   ) \omega_2^2 +  (\nu_2^2 -\mu_2^2)    \omega_2^4  }{\omega _1  \left( \omega_1^2 -  \omega_2^2   \right)^2      \left(\mu _2 \nu_0-  \mu_0 \nu_2 \right)^2 }   = \frac{1}{2 \omega_1 (\omega_1^2 -\omega_2^2 )} ,  \\
     \left[ \hat{b}, \hat{b}^\dagger  \right]   &=&   \frac{\nu_0^2- \mu_0^2 + 2( \mu_0 \mu_2 -\nu_0 \nu_2   ) \omega_1^2 +  (\nu_2^2 -\mu_2^2)    \omega_1^4  }{\omega _2  \left( \omega_1^2 -  \omega_2^2   \right)^2      \left(\mu _2 \nu_0-  \mu_0 \nu_2 \right)^2 }   =    \frac{1}{2 \omega_2 (\omega_2^2 -\omega_1^2 )}    .
\end{eqnarray}
In the last step we used (\ref{234}) and (\ref{dege}) to replace
\begin{equation}
	\Omega \rightarrow \frac{1}{8} \left(  \rho -\omega _1^2-\omega _2^2\right),  \,\,\,
	\nu^2  \rightarrow   \frac{1}{8} \left( \rho +\omega _1^2+\omega _2^2\right)  ,
\end{equation}
where $\rho =\sqrt{16 g^2+\left(\omega _1^2-\omega _2^2\right)^2}$. We have excluded here the choice with a minus sign in front of the square root, as in that case $\Gamma$ always maps to complex solutions. Using these creation and annihilation operators $\hat{a},\hat{a}^\dagger ,\hat{b}, \hat{b}^\dagger $, where we still need to identify which one is which, the second quantised version of our Hamiltonian (\ref{Hghost}) becomes 
\begin{equation}
    \hat{\cal H} = 2  \left( \omega_1^2 -  \omega_2^2   \right)  \left( \omega_1^2  \hat{a}^\dagger  \hat{a}  -\omega_2^2  \hat{b}^\dagger  \hat{b}  \right)  +\frac{1}{2} ( \omega_1 +  \omega_2 )    .     \label{Hv1}
\end{equation}
In this formulation the observed level crossing in the energy spectrum is evident as the model is simply build from two commuting copies of $SU(1,1)$ algebras. Next we construct the bosonic Fock space for this Hamiltonian.

\subsection{The ground state}
As candidates for the ground state we take again $\psi_0^{\epsilon,\eta}$ from (\ref{wansatz}), with $\epsilon,\eta$ labelling the four solution (\ref{abcd}) for $\alpha,\beta,\gamma$. Translating from the variables $\Omega, \nu$ to $\omega_1, \omega_2$  these solutions become
\begin{equation}
	\alpha_\epsilon^\eta = \frac{\rho + \Omega_{12}^\epsilon }{4 \eta  \sqrt{\Omega_{12}^\epsilon}}, \qquad
	\beta_\epsilon^\eta =  \frac{\rho -\Omega_{12}^\epsilon }{4 \eta  \sqrt{ \Omega_{12}^\epsilon }}, \qquad
	\gamma_\epsilon^\eta = -\frac{g}{\eta  \sqrt{ \Omega_{12}^\epsilon   }},
\end{equation}
with $\Omega_{12}^\epsilon =  \omega _1^2+\omega _2^2+2 \epsilon  \sqrt{\omega _1^2 \omega _2^2} $. Next we identify which operator annihilated the state $\psi_0^{\epsilon,\eta}$ depending on the choices for $\epsilon$, $\eta$ and the regimes. We find
\begin{align}
	\hat{a} \psi_0^{-+} &= 0,  \,\,\,\, \quad  \hat{a}^\dagger \psi_0^{--} = 0,&  &\text{for}  \,\,\, \omega_1, \omega_2 \in \{ \text{II}, \text{III},  \text{VI},  \text{VII}      \} , \label{annorm1} \\
	\hat{a}^\dagger \psi_0^{-+} &= 0,  \,\,\,\,\,\,\,  \quad  \hat{a} \psi_0^{--} = 0,&  &\text{for}    \,\,\,  \omega_1, \omega_2 \in \{ \text{I}, \text{IV},  \text{V},  \text{VIII}        \} ,    \\
	\hat{b} \psi_0^{-+} &= 0,  \,\,\,\,\,\,    \quad \hat{b}^\dagger \psi_0^{--}  = 0,&    &\text{for}  \,\,\,  \omega_1, \omega_2 \in \{ \text{I}, \text{III},  \text{V},  \text{VII}        \} ,  \\
	\hat{b}^\dagger \psi_0^{-+} &= 0,  \,\,\,\,\,\,\,   \quad  \hat{b} \psi_0^{--} = 0,&  & \text{for}    \,\,\,  \omega_1, \omega_2 \in \{ \text{II}, \text{IV},  \text{VI},  \text{VIII}       \},     \label{annorm4}
\end{align}
and 
\begin{align}
	\hat{a} \psi_0^{++} &= 0,  \,\,\,\, \quad  \hat{a}^\dagger \psi_0^{+-} = 0,&  &\text{for}  \,\,\, \omega_1, \omega_2 \in \{ \text{I}, \text{II},  \text{VII},  \text{VIII}      \} ,  \\
	\hat{a}^\dagger \psi_0^{++} &= 0,  \,\,\,\,\,\,\,  \quad  \hat{a} \psi_0^{+-} = 0,&  &\text{for}    \,\,\,  \omega_1, \omega_2 \in \{ \text{III}, \text{IV},  \text{V},  \text{VI}        \} ,    \\
	\hat{b} \psi_0^{++} &= 0,  \,\,\,\,\,  \quad  \hat{b}^\dagger \psi_0^{+-}  = 0,   &    &\text{for}  \,\,\,  \omega_1, \omega_2 \in \{ \text{I}, \text{II},  \text{III},  \text{IV}        \} ,  \\
	\hat{b}^\dagger \psi_0^{++} &= 0,   \,\,\,\,\,\,\,  \quad \hat{b} \psi_0^{+-} = 0  &  & \text{for}    \,\,\,  \omega_1, \omega_2 \in \{ \text{V}, \text{VI},  \text{VII},  \text{VIII}       \}.
\end{align}
Notice that according to our table 1 only $\psi_0^{-+} $ are candidates posses non-empty regimes where all three desirable properties are satisfied.

In order to identify the ground state energy it is most convenient to bring all the annihilation operators to the right in the expression for the Hamiltonian. Thus it is useful to re-write the Hamiltonian in some slightly alternative forms by using the commutation relations (\ref{comm1}) and (\ref{comm2}). We easily obtain  
\begin{eqnarray}
	\hat{\cal H}& =& 2  \left( \omega_1^2 -  \omega_2^2   \right)  \left( \omega_1^2  \hat{a}  \hat{a}^\dagger   -\omega_2^2  \hat{b}^\dagger  \hat{b}  \right)  +\frac{1}{2} ( \omega_2 -  \omega_1 ) ,  \label{Hv2} \\
	& =& 2  \left( \omega_1^2 -  \omega_2^2   \right)  \left( \omega_1^2   \hat{a}^\dagger  \hat{a}  -\omega_2^2    \hat{b}   \hat{b}^\dagger \right)  +\frac{1}{2} ( \omega_1 -  \omega_2 ) ,    \label{Hv3} \\
	& =& 2  \left( \omega_1^2 -  \omega_2^2   \right)  \left( \omega_1^2  \hat{a}  \hat{a}^\dagger   -\omega_2^2   \hat{b}  \hat{b}^\dagger   \right)  - \frac{1}{2} ( \omega_1 +  \omega_2 ) .  \label{Hv4}
\end{eqnarray}
Using the relations (\ref{annorm1})-(\ref{annorm4}) to identify the version with all annihilation operators to the right, we can directly read off the ground state energies in the different regimes as 
\begin{eqnarray}
	\text{III}, \text{VII} &:&  \,\,\,\,  \hat{\cal H} \psi_0^{-+}=\frac{1}{2} (\omega_1 +\omega_2) \psi_0^{-+}, \quad  \hat{\cal H} \psi_0^{--}=-\frac{1}{2} (\omega_1 +\omega_2)  \psi_0^{--} ,\\
		\text{I}, \text{V} &:&  \,\,\,\,  \hat{\cal H} \psi_0^{-+}=\frac{1}{2} (\omega_2 -\omega_1)  \psi_0^{-+} ,  \quad \hat{\cal H} \psi_0^{--}=\frac{1}{2} (\omega_1 -\omega_2)  \psi_0^{--} , \\
			\text{II}, \text{VI} &:&  \,\,\,\,  \hat{\cal H} \psi_0^{-+}=\frac{1}{2} (\omega_1 -\omega_2)  \psi_0^{-+}, \quad \hat{\cal H} \psi_0^{--}=\frac{1}{2} (\omega_2 -\omega_1)   \psi_0^{--},   \\
				\text{IV}, \text{VIII} &:&  \,\,\,\,  \hat{\cal H} \psi_0^{-+}=-\frac{1}{2} (\omega_1 +\omega_2)  \psi_0^{-+}, \quad \hat{\cal H} \psi_0^{--}=\frac{1}{2} (\omega_1 +\omega_2)   \psi_0^{--} .
\end{eqnarray}
For the ground state $ \psi_0^{-+}$ we used version (\ref{Hv1}) in the regions III, VII, version  (\ref{Hv2}) in the regions I V,  (\ref{Hv3}) in the regions II, VI and  (\ref{Hv4}) in the regions IV, VIII. For $ \psi_0^{--}$ we used version (\ref{Hv1}) in the regions IV, VIII, version  (\ref{Hv2}) in the regions II VI,  (\ref{Hv3}) in the regions I, V and  (\ref{Hv4}) in the regions III, VII. When translating back to $\alpha, \beta, \gamma$ we find that in all cases we obtain $E_0 = \alpha -\beta$, i.e. we have $\hat{\cal H} \psi_0 = (\alpha -\beta) \psi_0  $ in all cases in the respective regimes.

\subsection{The energy spectrum and the excited states}
The eigensystem can be defined in the usual way by building up a Fock space from our sets of creation and annihilation operators. Since we observed in the previous section that we obtain identical expressions when we convert the frequency representation back to the original model parameters, it suffices to consider one regime. For definiteness we take here the regime I. The exited states are defined by applying successive powers of the two creation operators,  $\hat{a}$ and   $ \hat{b}^\dagger$ in this case, on the ground state. The normalised eigenfunctions of the exited states are then
\begin{equation}
    \bar{ \psi}_{n,m} : = \frac{1}{ \sqrt{N_{n,m}}} \hat{a}^n  \left(  \hat{b}^\dagger \right)^m\psi_0^{-+}, \quad N_{n,m} = \frac{2 \pi }{\sqrt{n! m! \omega_1 \omega_2 } }  \frac{1}{ 2^{n+m} \omega_2^m \omega_1^n  (\omega_1^2 -\omega_2^2)^{n+m}  }.
\end{equation}
The corresponding eigenenergy spectrum is computed to
\begin{equation}
	\bar{E}_{n,m}  = \left(  \frac{1}{2}  -n       \right)   \omega_1 +   \left( m-  \frac{1}{2}         \right)   \omega_2,   \qquad  n,m \in \mathbb{N}_0 .
\end{equation}
We observe the typical feature in HTDT of the spectrum being unbounded from below for normalisable states, as in the regime I both frequencies are positive.
For the non normalisable states we may, however, construct a bounded system. For instance, in the regime I we find
\begin{equation}
       \hat{\cal H} \psi_{n,m}^{++} = \left[ \frac{1}{2} (n+1) \omega_1 + \frac{1}{2} (m+1) \omega_2,    \right]  \psi_{n,m}^{++}, \qquad n,m \in \mathbb{N}_0 ,
\end{equation}
which clearly corresponds to a spectrum bounded from below since both frequencies are positive in the regime I. Evidently, both cases suffer from undesirable features, but given the choice between normalisable wavefunction and unbounded spectra from below versus non-normalisable wavefunctions and bounded spectra, the latter option seems to be less attractive.

Finally, we directly compare the eigensystem obtained here with the solution found from the recurrence relations in the previous section by transforming the $\alpha, \beta,\gamma$ dependence into an $\omega_1,\omega_2$ dependence. Starting with the energy spectrum from (\ref{eab1}), (\ref{eab2}) we obtain in the regime I
\begin{eqnarray}
	E_{Nn}^+ &=& \left(\frac{1}{2}-n\right) \omega _1 +  \left(N-n+\frac{3}{2}\right)  \omega _2= \bar{E}_{n-1,1+N-n} , \,\, n=1, \ldots , \left\lfloor \frac{N}{2}\right\rfloor, \, \\
	E_{Nn}^- &=&  \left(n-N-\frac{3}{2}\right)  \omega _1  +\left(n-\frac{1}{2}\right) \omega _2  =  \bar{E}_{1+N-n,n-1} , \\ 
	E_{NN}    &=&  -\left(\frac{N}{2}+\frac{1}{2}\right) \omega _1+\left(\frac{N}{2}+\frac{1}{2}\right) \omega _2 =  \bar{E}_{N/2,N/2},     \quad     (N+1 \bmod 2). \qquad \qquad
\end{eqnarray}
The same identification for the quantum numbers then holds for the wavefunctions. Ignoring overall factors as the wavefunctions for the solutions of the recurrence are not normalised yet, we identify
\begin{equation}
	 \psi^+_{N n} \propto \bar{\psi}_{n-1,1+N-n}, \qquad  \psi^-_{N n} \propto \bar{\psi}_{1+N-n,n-1}, \qquad  \psi_{N N } \propto \bar{\psi}_{N/2,N/2} .
\end{equation}
Thus, both approaches are consistent with each other and lead to the same result.

\section{Model properties}
\begin{figure}[h]
	\centering         
	\begin{minipage}[b]{0.24\textwidth}     
		\includegraphics[width=\textwidth]{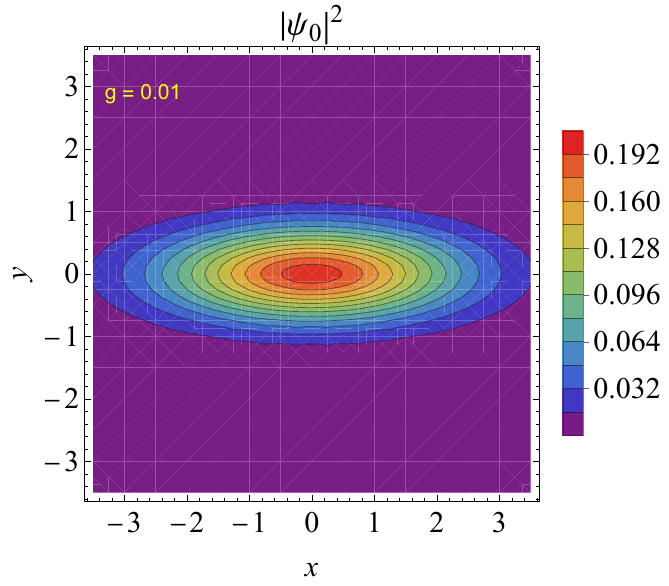}
	\end{minipage}   
	\begin{minipage}[b]{0.24\textwidth}           
		\includegraphics[width=\textwidth]{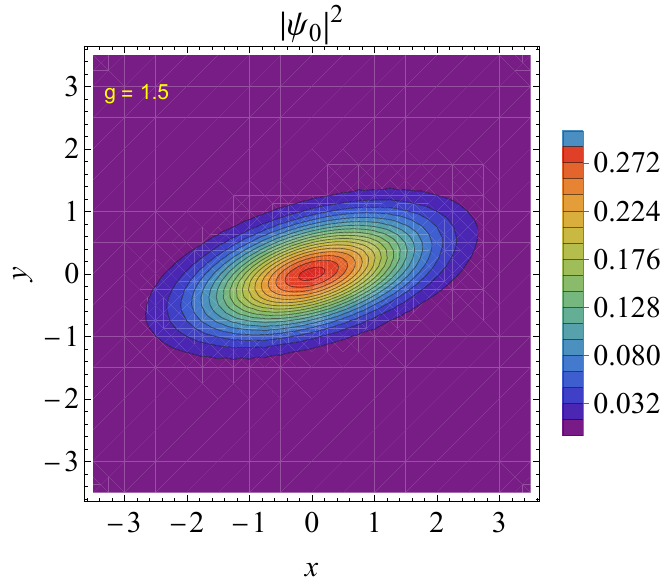}
	\end{minipage}    
	\begin{minipage}[b]{0.24\textwidth}    
		\includegraphics[width=\textwidth]{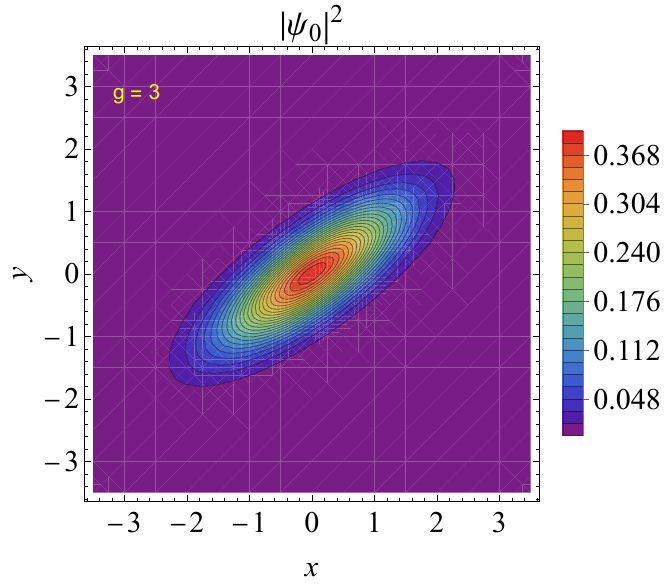}
	\end{minipage}   
	\begin{minipage}[b]{0.24\textwidth}           
		\includegraphics[width=\textwidth]{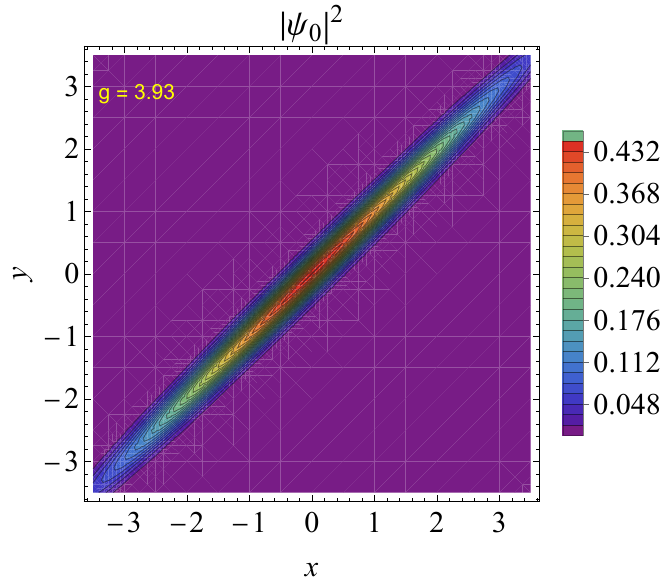}
	\end{minipage}    
	\begin{minipage}[b]{0.24\textwidth}     
		\includegraphics[width=\textwidth]{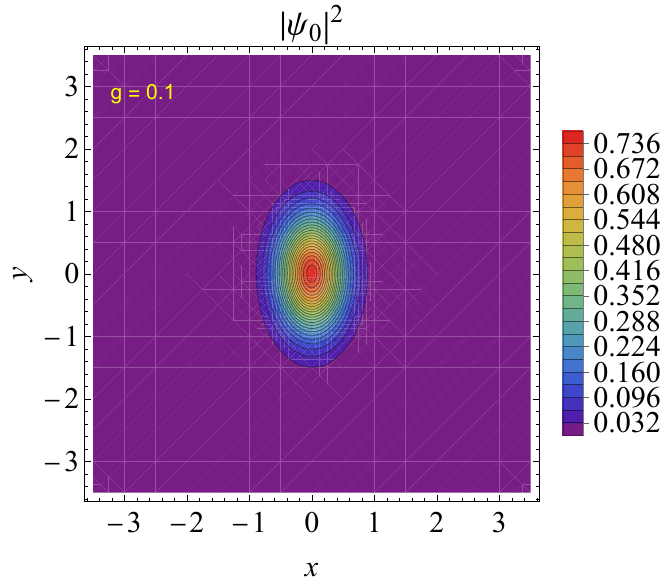}
	\end{minipage}   
	\begin{minipage}[b]{0.24\textwidth}           
		\includegraphics[width=\textwidth]{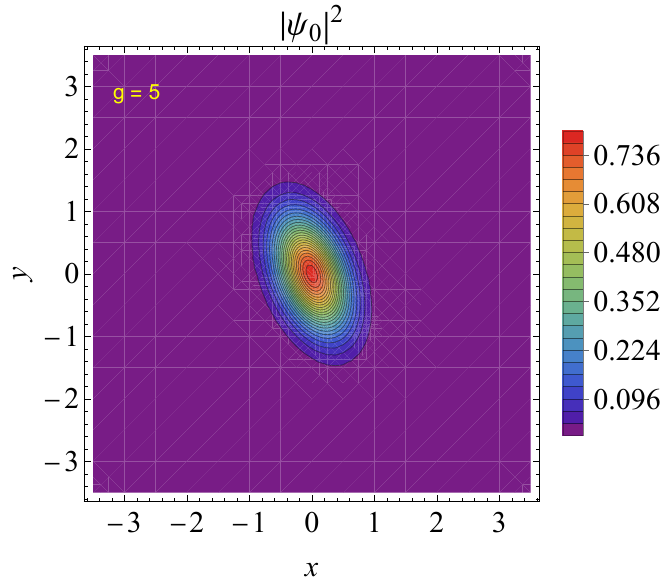}
	\end{minipage}    
	\begin{minipage}[b]{0.24\textwidth}    
		\includegraphics[width=\textwidth]{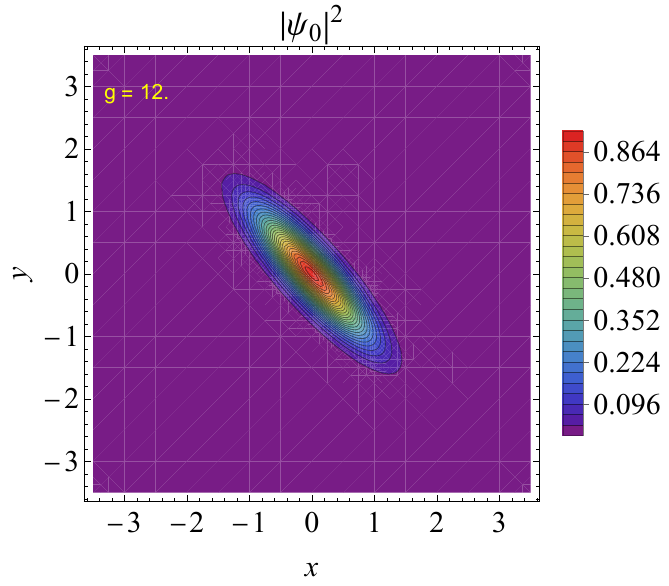}
	\end{minipage}   
	\begin{minipage}[b]{0.24\textwidth}           
		\includegraphics[width=\textwidth]{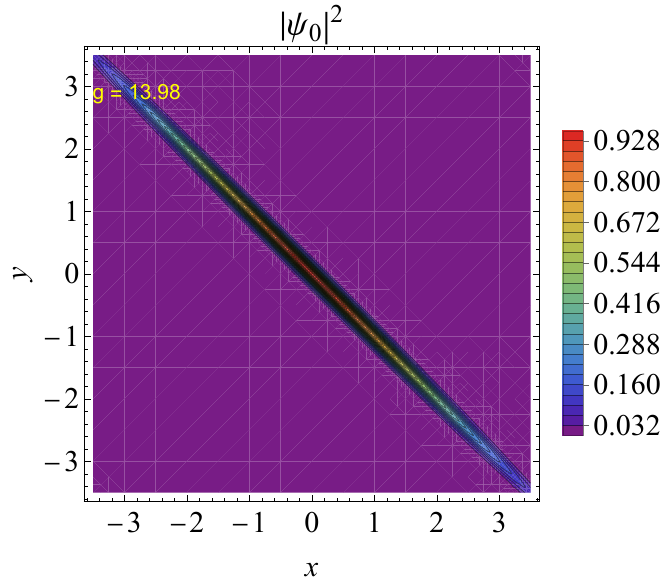}
	\end{minipage}    
	\caption{Probability densities for the ``ground state" for different coupling constants $g$ for the $\epsilon=\eta= -1$  branch with $\nu=4$, $\Omega = -2$ in the upper row and  $\epsilon=-\eta= -1$ branch with $\nu=0.2$, $\Omega = -4$ in the lower row. }
	\label{Prob}
\end{figure}

Having explicitly solved the quantum system, we are now in a position to calculate all relevant physical properties of our ghost model. We begin by computing the probability distribution $\vert  \psi(x,y) \vert^2$ in $\mathbb{R}^2$. Our results for the normalisable wavefunctions are shown in figure \ref{Prob}. 

For a small coupling constant $g$, we observe that the ground state is most localized at the origin, with the probability density falling off symmetrically in the negative and positive $x$ and $y$ directions, depending on the strength of the potential in each direction. As the coupling constant increases, the horizontal axis of the elliptical region in which the particle is localized begins to tilt. The ellipse becomes increasingly squeezed until its axis aligns with the $y=x$ or $y=-x$ diagonal in the $\epsilon=-\eta =-1$ or $\epsilon=\eta =-1$ case, respectively. Eventually the wavefunction collapses into a line when $g$ reaches the boundary of the defining domain in parameter space.

\begin{figure}[h]
	\centering         
	\begin{minipage}[b]{0.24\textwidth}     
		\includegraphics[width=\textwidth]{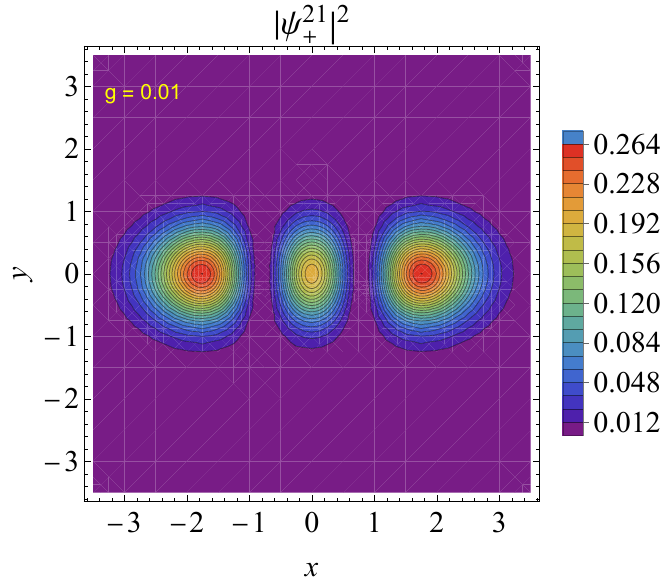}
	\end{minipage}   
	\begin{minipage}[b]{0.24\textwidth}           
		\includegraphics[width=\textwidth]{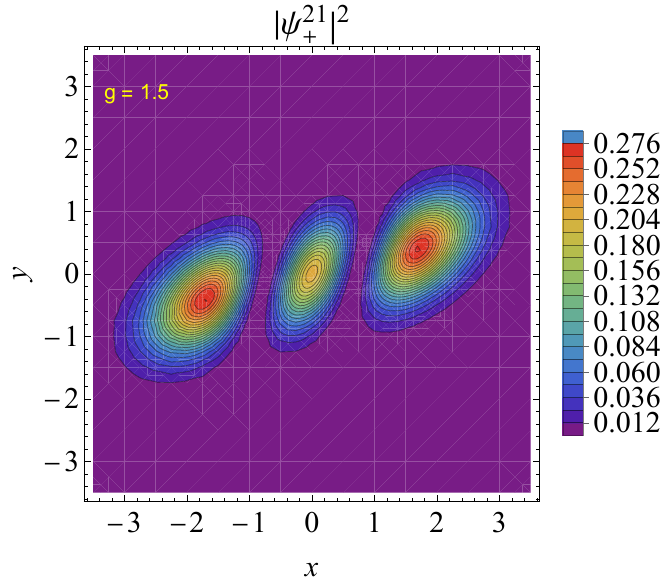}
	\end{minipage}    
	\begin{minipage}[b]{0.24\textwidth}    
		\includegraphics[width=\textwidth]{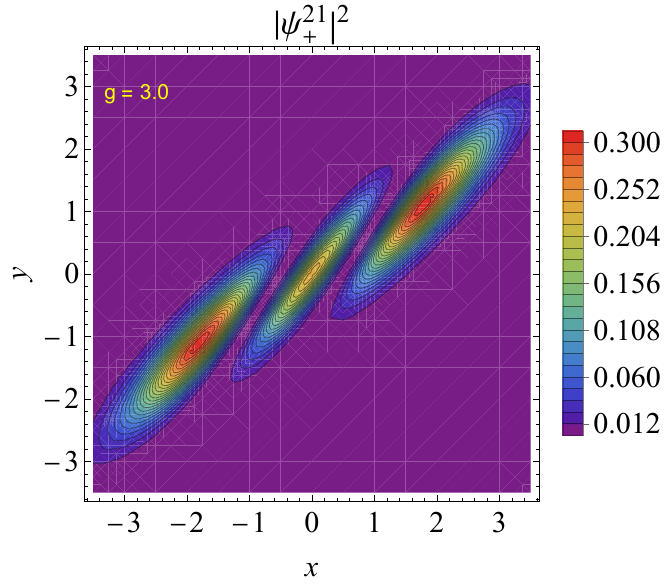}
	\end{minipage}   
	\begin{minipage}[b]{0.24\textwidth}           
		\includegraphics[width=\textwidth]{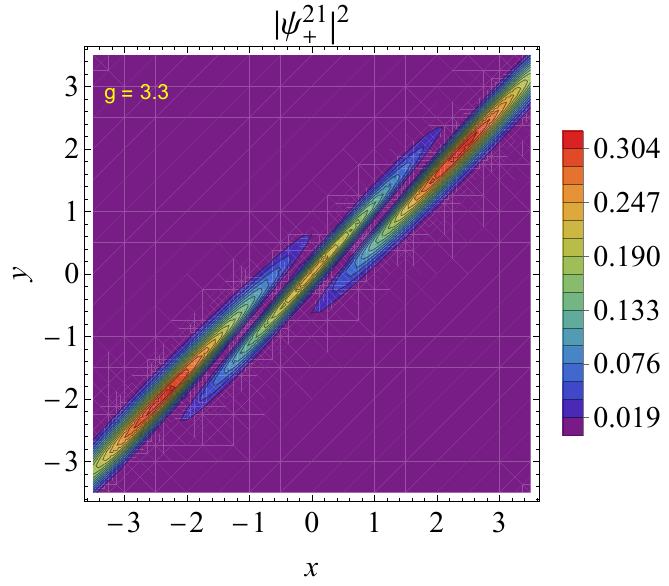}
	\end{minipage}    
	\caption{Probability densities for the   $\psi_{21}^+$ state for different coupling constants $g$ for the $\epsilon=\eta= -1$  branch with $\nu=4$, $\Omega = -2$.}
	\label{Prob1}
\end{figure}

For the exited states the behaviour is similar for each of the nodes of the wavefunctions as seen in figure \ref{Prob1} for an example.

\begin{figure}[h]
	\centering         
	\begin{minipage}[b]{0.5\textwidth}           \includegraphics[width=\textwidth]{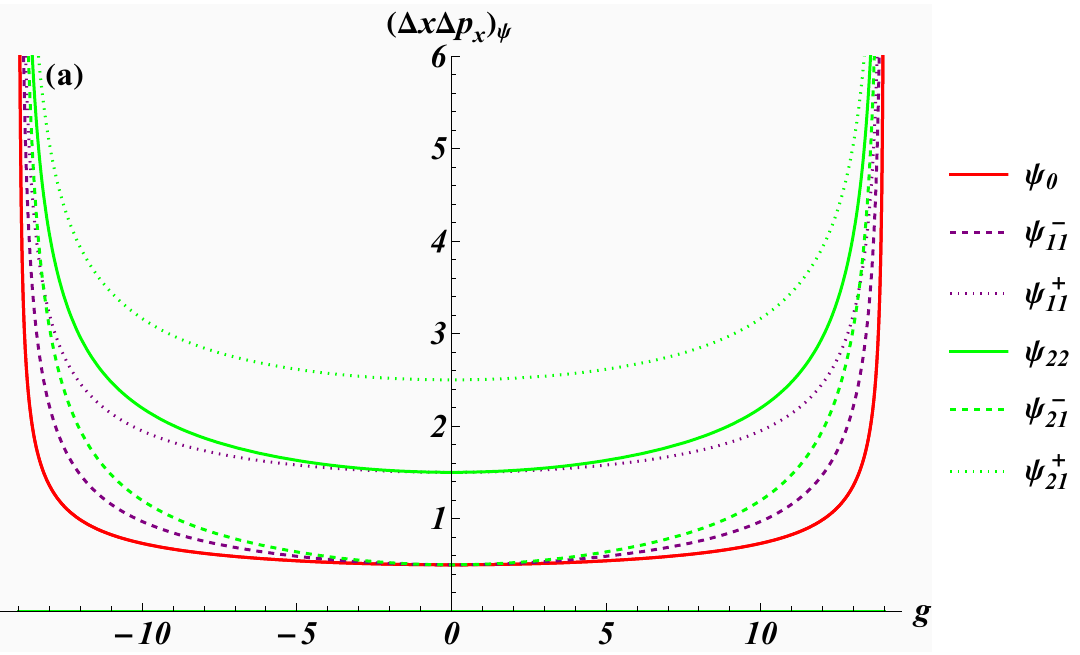}
	\end{minipage}   
	\begin{minipage}[b]{0.435\textwidth}           
	 \includegraphics[width=\textwidth]{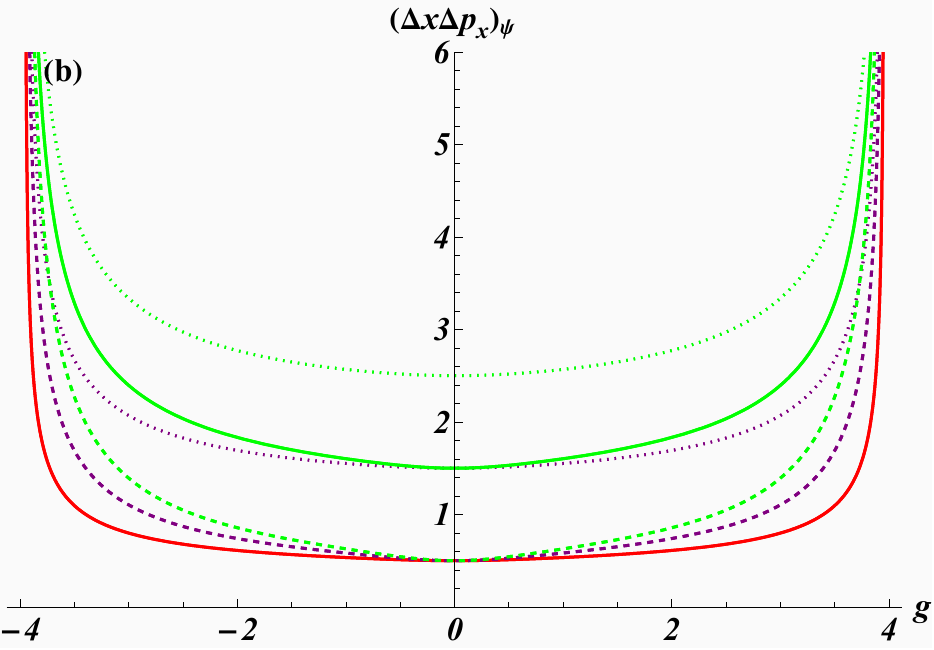}
	\end{minipage}    
	\caption{Uncertainty relations (\ref{un1})-(\ref{un4}) for varying coupling constant $g$ for the $\epsilon=\eta= -1$  branch with $\nu=4$, $\Omega = -2$ in panel (a) and  $\epsilon=-\eta= -1$ branch with $\nu=0.2$, $\Omega = -4$ in panel (b). }
	\label{uncertplot}
\end{figure}

Another interesting quantity to compute is the uncertainty relation 
\begin{equation}
	\left(  \Delta x \Delta p_x \right)_{\psi}  = \sqrt{  \langle  x^2   \rangle_\psi  -  \langle  x   \rangle_\psi ^2        }  \sqrt{  \langle  p_x^2   \rangle_\psi  -  \langle  p_x   \rangle_\psi ^2        }  = \sqrt{  \langle  x^2   \rangle_\psi          }  \sqrt{  \langle  p_x^2   \rangle_\psi      }  .  \label{uncee}
\end{equation}
Using the explicit solution we find for the first examples
\begin{eqnarray}
	\left(  \Delta x \Delta p_x \right)_{\psi_{11}  } &=&	\frac{1}{2} \sqrt{ \frac{\alpha \beta}{ \alpha \beta - \gamma^2 } }, \label{un1} \\
		\left(  \Delta x \Delta p_x \right)_{\psi_{22}  } &=&	\frac{3}{2} \sqrt{ \frac{\alpha \beta}{ \alpha \beta - \gamma^2 } },  \\
	\left(  \Delta x \Delta p_x \right)_{\psi_{11}^\pm}   &=&	\frac{1}{2} \sqrt{ 5 \pm \frac{4 (\alpha +\beta )}{\sqrt{(\alpha
				+\beta )^2-4 \gamma ^2}}     + \gamma ^2 \left(\frac{4}{(\alpha +\beta )^2-4 \gamma ^2}+\frac{3}{\alpha  \beta -\gamma ^2}\right)}, \\
		\left(  \Delta x \Delta p_x \right)_{\psi_{21}^\pm}   &=&	\frac{1}{2} \sqrt{ 13\pm \frac{12 (\alpha +\beta )}{\sqrt{(\alpha
				+\beta )^2-4 \gamma ^2}}     + \gamma ^2 \left(\frac{16}{(\alpha +\beta )^2-4 \gamma ^2}+\frac{5}{\alpha  \beta -\gamma ^2}\right)}. \label{un4} 
\end{eqnarray}
Similar expressions are obtained for the uncertainties in the $y$ direction. We depict these expressions in figure \ref{uncertplot}.

\begin{figure}[h]
	\centering         
	\begin{minipage}[b]{0.24\textwidth}     
		\includegraphics[width=\textwidth]{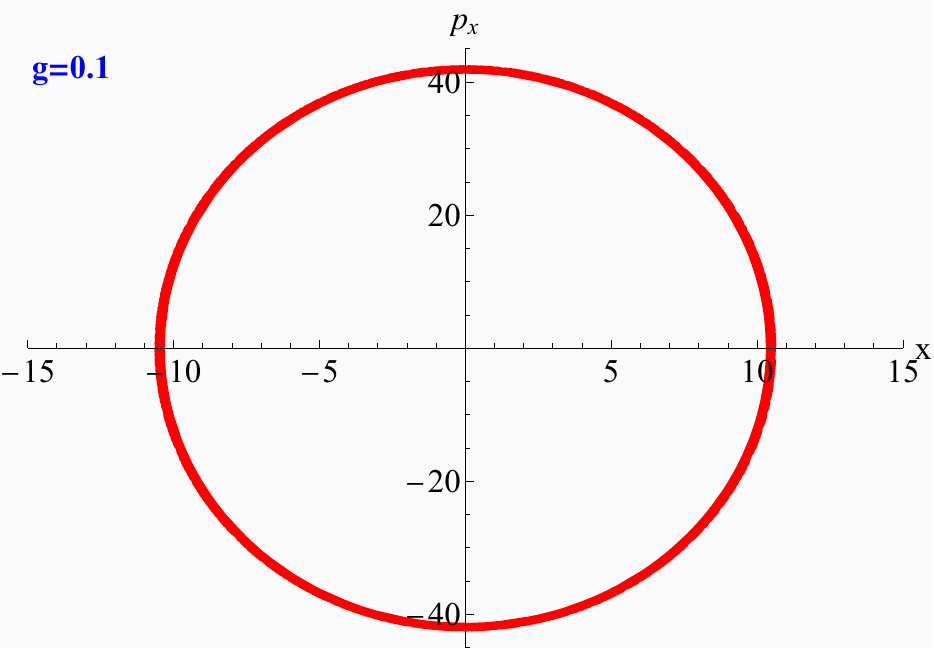}
	\end{minipage}   
	\begin{minipage}[b]{0.24\textwidth}           
		\includegraphics[width=\textwidth]{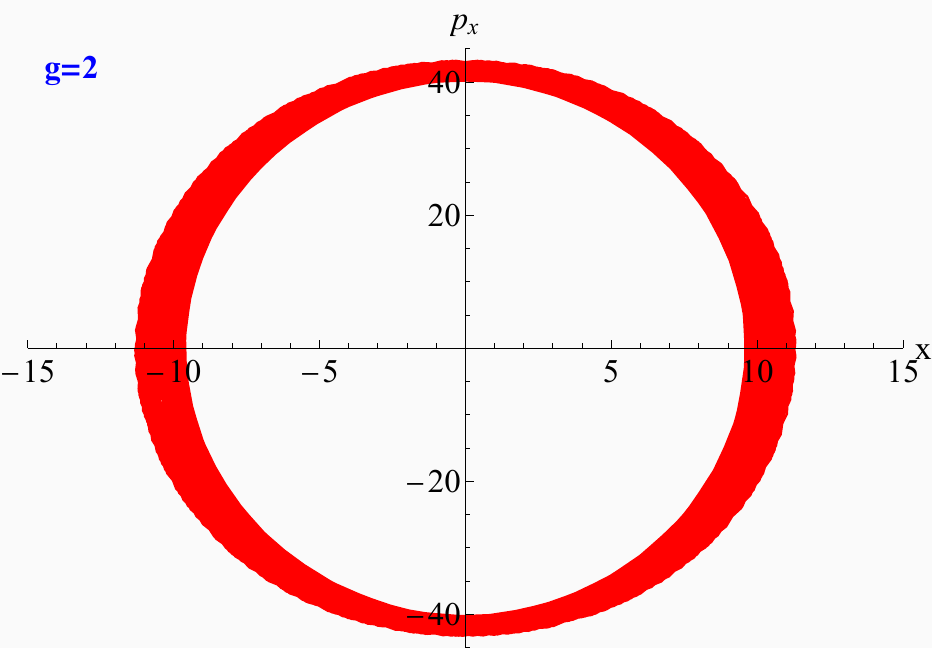}
	\end{minipage}    
	\begin{minipage}[b]{0.24\textwidth}    
		\includegraphics[width=\textwidth]{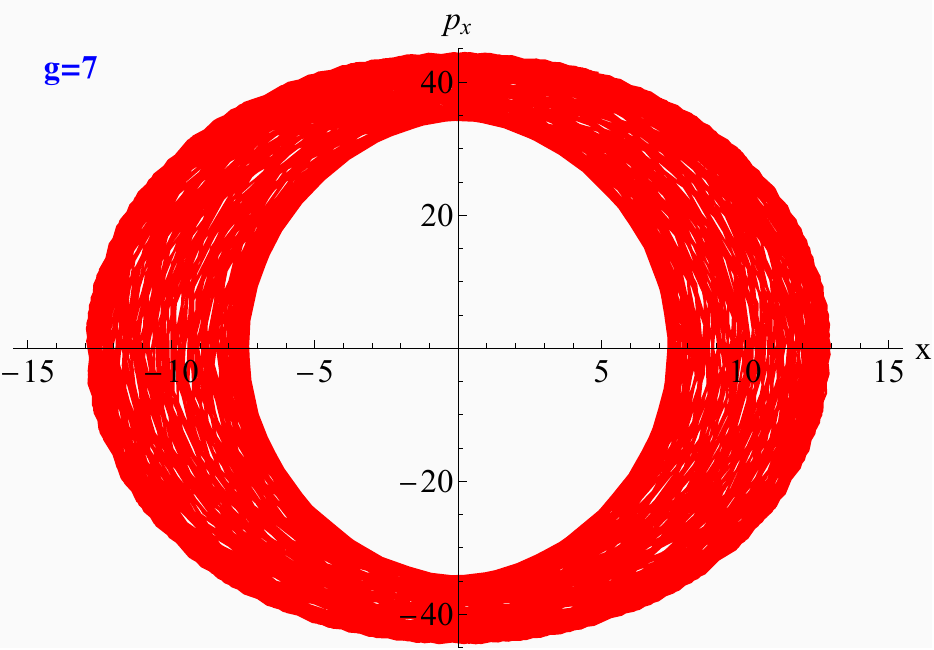}
	\end{minipage}   
	\begin{minipage}[b]{0.24\textwidth}           
		\includegraphics[width=\textwidth]{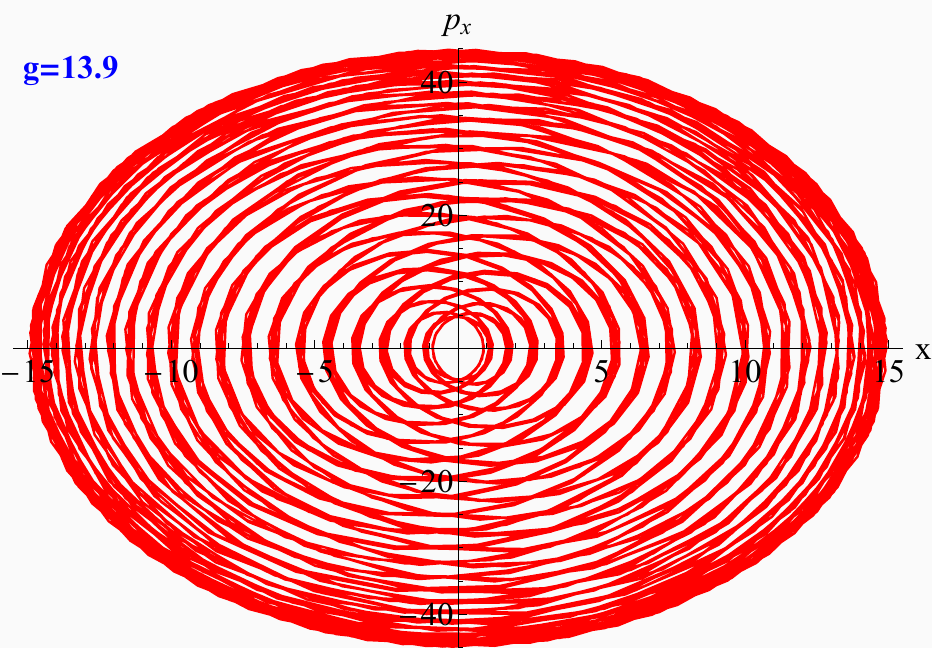}
	\end{minipage}    
	\caption{Classical trajectories from (\ref{clsol1})-(\ref{clsol4}) in the $(x,p_x)$ phase space in regime I with $\Omega =-1$, $\nu=4$.}
	\label{Classical}
\end{figure}

Reassuringly, we note first that the fundamental bound of $1/2$ is respected by all solutions and even saturated for small $g$ for $\psi_0$,   $\psi_{11}^-$ and $\psi_{22}^-$. Furthermore, we observe that when the coupling constant approaches the boundary of the domain in the parameter regime the uncertainties tend to infinity. This behaviour is qualitatively reproduced in the classical phase space. In figure \ref{Classical} we plot the classical solutions (\ref{clsol1})-(\ref{clsol4}) with constants $a=b=0.5$ from $t=0$ to some arbitrary large time $t=150$ for the regime I. We observe that for small values of the coupling constant the solutions are well localised in the $(x,p_x)$-phase space. This localisation becomes more and more fuzzy when the coupling constant increases and eventually a whole region is filled out when the boundary of the defining domain is approached. When the cut-off time becomes very large the phase space is also filled for small values of $g$, but it takes considerable longer to achieve that.

\section{Conclusions}

In this work, we investigated an exactly solvable ghostly quantum system and explored its quantisation through two distinct approaches. By explicitly mapping our system onto the Pais-Uhlenbeck oscillator via a symplectic transformation, we established a direct correspondence between their dynamical structures in certain parameter regimes, thereby demonstrating that our model can be interpreted as a higher time-derivative theory (HTDT) in disguise. This correspondence allowed us to quantise the system in two alternative ways where the transformation map is well-defined.

We obtained exact solutions to the Schrödinger equation using an Ansatz that led to theories defined in four different superselection sectors. In each of them we were led to a set of coupled three-term recurrence relations, which we solved systematically to construct normalisable wavefunctions and their associated energy spectra. Our findings were further validated by a second-quantisation approach, where we built a bosonic Fock space representation of the model. Both methods yielded consistently compatible results, confirming the reliability of our analytical solutions.

One of the insights from our study is the intricate balance between normalisability and spectral boundedness in ghostly quantum systems similar to the features found in HTDT. While the sectors with normalisable solutions exhibit unbounded spectra, the two sectors with non-normalisable wavefunctions can lead to bounded spectra either from above or below.
The $\cal{PT}$-symmetry of the system is most relevant in the two latter cases in which it explains the occurrence of exceptional points and a spontaneously broken regime with pairs of complex conjugate eigenvalues.

The model properties in the low coupling regime revealed an interesting behaviour with the coupling constant approaching the boundary of the defining parameter regime. Our analysis of probability densities and uncertainty relations illustrated how the coupling parameter influences the localisation properties of the system, with increasing interactions leading to pronounced quantum fluctuations and loss of phase space localisation. 

At this stage we did not provide any analysis about the model's behaviour at the point in parameter space that is equivalent to the degenerate point in the PU-model, which is known to be problematic already at the classical level with its well-known divergent solutions and so far no widely accepted quantised version. Similarly it is unclear whether our ghostly model possess consistent solutions in the strong coupling regime.

Our results contribute to the broader understanding of higher-derivative quantum models and ghostly excitations in quantum mechanics, providing a novel exactly solvable framework for further explorations.
 Future directions include extending this approach to field-theoretic models, investigating interactions with external potentials, and long term exploring potential applications to quantum gravity scenarios where higher-derivative corrections play a crucial role.
 
 \medskip

\noindent {\bf Acknowledgments}: BT is supported by a City St George's, University of London Research Fellowship.

\newif\ifabfull\abfulltrue


\begin{thebibliography}{10}
	
	\bibitem{ostrogradsky1850memoire}
	M.~Ostrogradsky,
	\newblock {\em M{\'e}moire sur les {\'e}quations diff{\'e}rentielles relatives
		an probl{\'e}me des isop{\'e}rim{\'e}tres}, volume VI 4,
	\newblock 1850.
	
	\bibitem{bateman1931diss}
	H.~Bateman,
	\newblock On dissipative systems and related variational principles,
	\newblock Phys. Rev. {\bf 38}(4), 815 (1931).
	
	\bibitem{dekker1981clas}
	H.~Dekker,
	\newblock Classical and quantum mechanics of the damped harmonic oscillator,
	\newblock Phys. Rep. {\bf 80}(1), 1--110 (1981).
	
	\bibitem{pais1950field}
	A.~Pais and G.~E. Uhlenbeck,
	\newblock On field theories with non-localized action,
	\newblock Phys. Rev. {\bf 79}(1), 145 (1950).
	
	\bibitem{diez2024foundations}
	V.~E. D{\'\i}ez, J.~G. Gaset~Rif{\`a}, and G.~Staudt,
	\newblock Foundations of Ghost Stability,
	\newblock Fortschritte der Physik , 2400268 (2024).
	
	\bibitem{stelle77ren}
	K.~S. Stelle,
	\newblock Renormalization of higher-derivative quantum gravity,
	\newblock Phys. Rev. D {\bf 16}(4), 953 (1977).
	
	\bibitem{grav1}
	A.~A. Starobinsky,
	\newblock A new type of isotropic cosmological models without singularity,
	\newblock Phys. Lett. B {\bf 91}(1), 99--102 (1980).
	
	\bibitem{grav2}
	S.~L. Adler,
	\newblock Einstein gravity as a symmetry-breaking effect in quantum field
	theory,
	\newblock Rev. Mod. Phys. {\bf 54}(3), 729 (1982).
	
	\bibitem{grav3}
	A.~V. Smilga,
	\newblock Spontaneous generation of the Newton constant in the renormalizable
	gravity theory,
	\newblock ITEP preprint 63 (1982) 8 pp, arXiv preprint arXiv:1406.5613 (2014)
	(1982).
	
	\bibitem{modesto16super}
	L.~Modesto and I.~L. Shapiro,
	\newblock Superrenormalizable quantum gravity with complex ghosts,
	\newblock Phys. Lett. B {\bf 755}, 279--284 (2016).
	
	\bibitem{Hawking}
	S.~W. Hawking and T.~Hertog,
	\newblock Living with ghosts,
	\newblock Phys. Rev. D {\bf 65}(10), 103515 (2002).
	
	\bibitem{biswas2010towards}
	T.~Biswas, T.~Koivisto, and A.~Mazumdar,
	\newblock Towards a resolution of the cosmological singularity in non-local
	higher derivative theories of gravity,
	\newblock JCAP {\bf 2010}(11), 008 (2010).
	
	\bibitem{Salvio2}
	A.~Salvio,
	\newblock Dimensional transmutation in gravity and cosmology,
	\newblock Int. J. Mod. Phys. A {\bf 36}(08n09), 2130006	(2021).
	
	\bibitem{Salvio3}
	A.~Salvio,
	\newblock Quasi-conformal models and the early universe,
	\newblock EPJ C {\bf 79}(9), 750 (2019).
	
	\bibitem{Salvio4}
	A.~Salvio,
	\newblock A non-perturbative and background-independent formulation of
	quadratic gravity,
	\newblock JCAP {\bf 2024}(07), 092
	(2024).
	
	\bibitem{weldon98finite}
	H.~A. Weldon,
	\newblock Finite-temperature retarded and advanced fields,
	\newblock Nucl. Phys. B {\bf 534}(1-2), 467--490 (1998).
	
	\bibitem{mignemi1992black}
	S.~Mignemi and D.~L. Wiltshire,
	\newblock Black holes in higher-derivative gravity theories,
	\newblock Phys. Rev. D {\bf 46}(4), 1475 (1992).
	
	\bibitem{plyush89mass}
	M.~S. Plyushchay,
	\newblock Massless point particle with rigidity,
	\newblock Mod. Phys. Lett. A {\bf 4}(09), 837--847 (1989).
	
	\bibitem{Mpl}
	M.~S. Plyushchay,
	\newblock Massless particle with rigidity as a model for the description of
	bosons and fermions,
	\newblock Phys. Lett. B {\bf 243}(4), 383--388 (1990).
	
	\bibitem{dine1997comments}
	M.~Dine and N.~Seiberg,
	\newblock Comments on higher derivative operators in some SUSY field theories,
	\newblock Phys. Lett. B {\bf 409}(1-4), 239--244 (1997).
	
	\bibitem{smilga17ultrav}
	A.~Smilga,
	\newblock Ultraviolet divergences in non-renormalizable supersymmetric
	theories,
	\newblock Phys. of Part. and Nucl. Lett. {\bf 14}, 245--260 (2017).
	
	\bibitem{ghostconst}
	T.-J. Chen, M.~Fasiello, E.~A. Lim, and A.~J. Tolley,
	\newblock Higher derivative theories with constraints: Exorcising
	Ostrogradski's Ghost,
	\newblock J. Cos. Astro. Phys. {\bf 2013}(02), 042 (2013).
	
	\bibitem{salvio16quant}
	A.~Salvio and A.~Strumia,
	\newblock Quantum mechanics of 4-derivative theories,
	\newblock The EPJ C {\bf 76}, 1--15 (2016).
	
	\bibitem{fakeons}
	D.~Anselmi,
	\newblock Fakeons and Lee-Wick models,
	\newblock JHEP {\bf 2018}(2) (2018).
	
	\bibitem{bender2008no}
	C.~M. Bender and P.~D. Mannheim,
	\newblock No-ghost theorem for the fourth-order derivative Pais-Uhlenbeck
	oscillator model,
	\newblock Phys. Rev. Lett. {\bf 100}(11), 110402 (2008).
	
	\bibitem{raidal2017quantisation}
	M.~Raidal and H.~Veerm{\"a}e,
	\newblock On the quantisation of complex higher derivative theories and
	avoiding the Ostrogradsky ghost,
	\newblock Nucl. Phys. B {\bf 916}, 607--626 (2017).
	
	\bibitem{rivelles2003triviality}
	V.~O. Rivelles,
	\newblock Triviality of higher derivative theories,
	\newblock Phys. Lett. B {\bf 577}(3-4), 137--142 (2003).
	
	\bibitem{Kap1}
	D.~S. Kaparulin, S.~L. Lyakhovich, and A.~A. Sharapov,
	\newblock BRST analysis of general mechanical systems,
	\newblock J. of Geo. and Phys. {\bf 74}, 164--184 (2013).
	
	\bibitem{mandal2022bfv}
	B.~P. Mandal, V.~K. Pandey, and R.~Thibes,
	\newblock BFV quantization and BRST symmetries of the gauge invariant
	fourth-order Pais-Uhlenbeck oscillator,
	\newblock Nucl. Phys. B {\bf 982}, 115905 (2022).
	
	\bibitem{mann2005dirac}
	P.~D. Mannheim and A.~Davidson,
	\newblock Dirac quantization of the Pais-Uhlenbeck fourth order oscillator,
	\newblock Phys. Rev. A {\bf 71}(4), 042110 (2005).
	
	\bibitem{dama2006}
	E.~Damaskinsky and M.~Sokolov,
	\newblock Remarks on quantization of Pais--Uhlenbeck oscillators,
	\newblock J. Phys. A: Math. Gen. {\bf 39}(33), 10499
	(2006).
	
	\bibitem{bolonek2006comment}
	K.~Bolonek and P.~Kosinski,
	\newblock Comment on ``Dirac Quantization of Pais-Uhlenbeck Fourth Order
	Oscillator",
	\newblock arXiv preprint quant-ph/0612009  (2006).
	
	\bibitem{smilga2006ghost}
	A.~V. Smilga,
	\newblock Ghost-free higher-derivative theory,
	\newblock Phys. Lett. B {\bf 632}(2-3), 433--438 (2006).
	
	\bibitem{smilga2009comments}
	A.~V. Smilga,
	\newblock Comments on the dynamics of the Pais-Uhlenbeck oscillator,
	\newblock SIGMA. Symmetry, Integrability and Geometry: Methods and Applications
	{\bf 5}, 017 (2009).
	
	\bibitem{berra2015def}
	J.~Berra-Montiel, A.~Molgado, and E.~Rojas,
	\newblock Deformation quantization of the Pais--Uhlenbeck fourth order
	oscillator,
	\newblock Ann. Phys. {\bf 362}, 298--310 (2015).
	
	\bibitem{cumsille2016polymer}
	P.~Cumsille, C.~M. Reyes, S.~Ossandon, and C.~Reyes,
	\newblock Polymer quantization, stability and higher-order time derivative
	terms,
	\newblock Int. J. Mod. Phys. A {\bf 31}(09), 1650040
	(2016).
	
	\bibitem{mosta2011im}
	A.~Mostafazadeh,
	\newblock Imaginary-scaling versus indefinite-metric quantization of the
	Pais-Uhlenbeck oscillator,
	\newblock Phys. Rev. D 	{\bf 84}(10), 105018 (2011).
	
	\bibitem{Urries}
	F.~J. De~Urries and J.~Julve,
	\newblock Ostrogradski formalism for higher-derivative scalar field theories,
	\newblock J. Phys. A: Math. and Gen. {\bf 31}(33), 6949 (1998).
	
	\bibitem{weldon03quant}
	H.~A. Weldon,
	\newblock Quantization of higher-derivative field theories,
	\newblock Ann. Phys. {\bf 305}(2), 137--150 (2003).
	
	\bibitem{fring2024higher}
	A.~Fring, T.~Taira, and B.~Turner,
	\newblock Higher Time-Derivative Theories from Space--Time Interchanged
	Integrable Field Theories,
	\newblock Universe {\bf 10}(5), 198 (2024).
	
	\bibitem{Bender:1995rh}
	C.~M. Bender and G.~V. Dunne,
	\newblock {Quasiexactly solvable systems and orthogonal polynomials},
	\newblock J. Math. Phys. {\bf 37}, 6--11 (1996).
	
	\bibitem{Finkel}
	F.~Finkel, A.~Gonzalez-Lopez, and M.~A. Rodriguez,
	\newblock {Quasiexactly solvable potentials on the line and orthogonal
		polynomials},
	\newblock J. Math. Phys. {\bf 37}, 3954--3972 (1996).
	
	\bibitem{Krajewska}
	A.~Krajewska, A.~Ushveridze, and Z.~Walczak,
	\newblock Bender--Dunne Orthogonal Polynomials General Theory,
	\newblock Mod. Phys. Lett. A {\bf 12}, 1131--1144 (1997).
	
	\bibitem{KhareM}
	A.~Khare and B.~P. Mandal,
	\newblock A PT-invariant potential with complex QES eigenvalues,
	\newblock Phys. Lett. A {\bf 272}, 53--56 (2000).
	
	\bibitem{E2Fring}
	A.~Fring,
	\newblock E2-quasi-exact solvability for non-Hermitian models,
	\newblock J. Phys. {\bf A48}, 145301(19) (2015).
	
	\bibitem{AndTom5}
	A.~Fring and T.~Frith,
	\newblock Quasi-exactly solvable quantum systems with explicitly time-dependent
	Hamiltonians,
	\newblock Phys. Lett. A {\bf 383}(2-3), 158--163 (2019).
	
	\bibitem{FFT}
	A.~Felski, A.~Fring, and B.~Turner,
	\newblock Equivalent representations and alternative formulations of higher
	time-derivative models,
	\newblock in preparation.
	
	\bibitem{felski2018analytic}
	A.~Felski and S.~Klevansky,
	\newblock Analytic eigenvalue structure of a coupled-oscillator system beyond
	the ground state,
	\newblock Phys. Rev. A {\bf 98}(1), 012127 (2018).
	
	\bibitem{usmani94inv}
	R.~A. Usmani,
	\newblock Inversion of a tridiagonal Jacobi matrix,
	\newblock Linear Algebra and its Applications {\bf 212}(213), 413--414 (1994).
	
	\bibitem{vNW}
	J.~von Neuman and E.~Wigner,
	\newblock {\"U}ber merkw{\"u}rdige diskrete Eigenwerte. {\"U}ber das Verhalten
	von Eigenwerten bei adiabatischen Prozessen,
	\newblock Zeit. der Physik {\bf 30}, 467--470 (1929).
	
	\bibitem{Kato}
	T.~Kato,
	\newblock Perturbation Theory for Linear Operators,
	\newblock (Springer, Berlin)  (1966).
	
	\bibitem{berry2004physics}
	M.~V. Berry,
	\newblock Physics of nonhermitian degeneracies,
	\newblock Czech. J. Phys. {\bf 54}(10), 1039--1047 (2004).
	
	\bibitem{miri2019exceptional}
	M.-A. Miri and A.~Al{\`u},
	\newblock Exceptional points in optics and photonics,
	\newblock Science {\bf 363}(6422), 7709 (2019).
	
	\bibitem{PTbook}
	C.~M. Bender, P.~E. Dorey, C.~Dunning, A.~Fring, D.~W. Hook, H.~F. Jones,
	S.~Kuzhel, G.~Levai, and R.~Tateo,
	\newblock PT Symmetry: In Quantum and Classical Physics,
	\newblock (World Scientific, Singapore)  (2019).
	
	\bibitem{Gonoskov}
	I.~Gonoskov,
	\newblock Closed-form solution of a general three-term recurrence relation,
	\newblock Advances in Difference Equations {\bf 2014}, 196(12) (2014).
	
\end{thebibliography}

\end{document}